\documentclass[aps,pra,showpacs,amsmath,amssymb,amsfonts,tightenlines,twocolumn,lengthcheck]{revtex4-1}
\usepackage{diagbox}
\usepackage{multirow}
\usepackage{amsmath}
\usepackage{amsfonts}
\usepackage{graphicx}
\usepackage{epsfig}
\usepackage{color}
\usepackage{txfonts}
\usepackage[colorlinks,citecolor=blue]{hyperref}

\begin{document}

\title{Single-photon scattering in a giant-molecule waveguide-QED system}
\author{Xian-Li Yin}
\affiliation{Key Laboratory of Low-Dimensional Quantum Structures and Quantum Control of Ministry of Education, Key Laboratory for Matter Microstructure and Function of Hunan Province, Department of Physics and Synergetic Innovation Center for Quantum Effects and Applications, Hunan Normal University, Changsha 410081, China}
\author{Yu-Hong Liu}
\affiliation{Key Laboratory of Low-Dimensional Quantum Structures and Quantum Control of Ministry of Education, Key Laboratory for Matter Microstructure and Function of Hunan Province, Department of Physics and Synergetic Innovation Center for Quantum Effects and Applications, Hunan Normal University, Changsha 410081, China}
\author{Jin-Feng Huang}
\email{Corresponding author: jfhuang@hunnu.edu.cn}
\affiliation{Key Laboratory of Low-Dimensional Quantum Structures and Quantum Control of Ministry of Education, Key Laboratory for Matter Microstructure and Function of Hunan Province, Department of Physics and Synergetic Innovation Center for Quantum Effects and Applications, Hunan Normal University, Changsha 410081, China}
\author{Jie-Qiao Liao}
\email{Corresponding author: jqliao@hunnu.edu.cn}
\affiliation{Key Laboratory of Low-Dimensional Quantum Structures and Quantum Control of Ministry of Education, Key Laboratory for Matter Microstructure and Function of Hunan Province, Department of Physics and Synergetic Innovation Center for Quantum Effects and Applications, Hunan Normal University, Changsha 410081, China}

\begin{abstract}
We study the coherent single-photon scattering in a one-dimensional waveguide coupled to a giant artificial molecule consisting of two coupled giant atoms. Since each giant atom couples to the waveguide via two coupling points, the couplings of the molecule with the waveguide have three different coupling configurations: the separated-, braided-, and nested-coupling cases. We obtain the exact expressions of the single-photon transmission and reflection amplitudes with the real-space approach.  It is found that the behavior of the scattering spectra depends on the phase shift between two neighboring coupling points, the coupling configuration, and the coupling between the two giant atoms. Concretely, we study the photon scattering in both the Markovian and non-Markovian regimes, in which the photon propagating time between two neighboring coupling points is neglected and considered, respectively. Under the Markovian limit, the asymmetric Fano line shapes in different coupling configurations of the giant-molecule waveguide-QED system can be obtained by choosing proper phase shift, and the transmission window can be adjusted by the coupling strength between the two giant atoms in these three coupling configurations. In particular, multiple reflection peaks and dips in these configurations are revived in the non-Markovian regime. This paper will pave the way for the study of controllable single-photon devices based on the giant-molecule waveguide-QED systems.
\end{abstract}

\date{\today}
\maketitle
\section{Introduction}
The finding, understanding, and applications of the physical effects induced by the quantum light-matter interactions are at the heart of the field of quantum optics~\cite{Tannoudji1992}. The quantum light-matter interactions can be well realized in many physical platforms, such as cavity quantum electrodynamics (QED) systems~\cite{KimblePS1998,Haroche2001,Walther2006}, circuit-QED systems~\cite{Wallraff2021,Blais2004,Schoelkopf2004}, and waveguide-QED systems~\cite{Roy2017,Gu2017,Sheremet2021}. Different from the cavity-QED systems, the waveguides usually support a continuum of modes, and hence the finite bandwidth of the fields can be relaxed~\cite{Roy2017}. Significant progress on both theoretical and experimental works has been made in the field of waveguide QED. In particular, various setups have been developed in waveguide-QED systems, such as various qubits coupled to transmission lines~\cite{Astafiev2010,Lool2013,Liu2017}, photonic crystal waveguides~\cite{Goban2014,Goban2015}, and coupled-resonator arrays~\cite{Hartmann2006,Greentree2006}. These systems provide a good candidate for studying few-photon scattering~\cite{Shen2005a,Shen2005b,Zhou2008,Tsoi2008,Shi2009,Liao2009,Tsoi2009,Longo2010,Liao2010a,Liao2010,Zheng2010,Fan2010,Roy2011a,Chen2011,Shi2011,Rephaeli2011,Baragiola2012,Pletyukhov2012,Liao2013,Huang2013,ZLiao2015,ZLiao2016,Roulet2016,See2017,Hurst2018,Ke2019}. Meanwhile, chiral quantum optics~\cite{Zoller2017} and bound-state physics~\cite{Fan2007,Zheng2010,Liao2010,Shi2016} have also been widely investigated in waveguide-QED systems. In addition, it has been found that the propagating photons in a one-dimensional waveguide can mediate the long-range interactions between remote atoms~\cite{Baranger2013}, which are critical for the realization of quantum networks~\cite{Kimble2008}.

In waveguide-QED systems, several methods have been developed to study few-photon scattering. These methods include the real-space Bethe-ansatz method~\cite{Shen2005a,Shen2005b,Tsoi2008}, the Lehmann-Symanzik-Zimmermann reduction~\cite{Shi2009,Shi2011}, the wave-packet evolution approach~\cite{Longo2010,Liao2010a,Liao2010,Chen2011,Liao2013,ZLiao2016}, the Lippmann-Schwinger formalism~\cite{Zheng2010,Roy2011a,Huang2013},  the input-output theory~\cite{Fan2010,Rephaeli2011},  the master equation method~\cite{Baragiola2012}, various diagrammatic techniques~\cite{Pletyukhov2012,Roulet2016,See2017}, as well as the Dyson series summation~\cite{Hurst2018}. In particular, the two-photon scattering has been analytically solved in a Kerr nonlinear cavity~\cite{Liao2010} and a cavity optomechanical system~\cite{Liao2013} with the wave-packet evolution approach.

It should be pointed out that the light-matter interactions have previously considered small atoms coupled to electromagnetic fields, where the atomic dimensions are much smaller than the wavelengths of the fields. Thus we can reasonably regard the atoms as point-like and utilize the dipole approximation~\cite{Wall2008} to deal with the light-atom interactions. Recently, the giant atoms~\cite{Kockum2020Rev}, as an emerging playground in quantum optics, have attracted great interest from many peers. Many interesting quantum optical effects induced by giant atoms have been predicted, including the frequency-dependent relation and the Lamb shift~\cite{Kockum2014}, decoherence-free subspace~\cite{Kockum2018,Ciccarello2020,Carollo2020,Kockum2022pra}, nonexponential decay~\cite{Guo2017,Andersson2019,Longhi2020,SGuo2020,2022prr},  creation of bound states~\cite{Guo2020,Wang2020,WangX2021}, and electromagnetically induced transparency~\cite{Ask2020,Zhao2021}. Up to now, the giant atoms have been experimentally realized in several physical platforms~\cite{Gustafsson2014,Manenti2017,Andersson2019,Delsing2020,Kannan2020,Vadiraj2021}, such as superconducting qubits coupled with either surface acoustic waves (SAWs) or open transmission lines.

Recently, much attention has been paid to the study of photon scattering in giant-atom waveguide-QED systems~\cite{Wang2020,Du2021OE,Du2021PRA,Du2021PRR,Cai2021,Jia2021,Xue2021,Xiang2022}. In these systems, giant atoms can be engineered to couple with the waveguide at multiple points with large separation distance, and hence the dipole approximation is no longer valid. In particular, the distance between different coupling points can be comparable to or much larger than the general wavelengths, which leads to many interesting phenomena induced by the quantum interference effect. Additionally, the decoherence-free interaction has been predicted~\cite{Kockum2018} and experimentally observed~\cite{Kannan2020}. In Ref.~\cite{Jia2021}, the single-photon scattering in a waveguide-QED system containing double giant atoms has been studied via the real-space method. In the multiple-scatter scheme, a natural and important question is how do the inner couplings between scatters affect the photon transport properties. Physically, the inner coupling provides new excitation transfer channels, and hence quantum interference will modulate the photon scattering process.

Motivated by the above question, in this paper we study single-photon scattering in the giant-artificial-molecule waveguide system, where the giant molecule is formed by two coupled giant atoms.  Both the exact transmission and reflection amplitudes are derived by using the real-space method. In the Markovian regime, we find the asymmetric Fano line shapes~\cite{Fano1961,Miroshnichenko2010} and the Rabi splitting-like phenomenon in this system. In particular, we investigate the phenomenon of the Fano resonance in different coupling configurations when the phase shift takes various values in detail. Moreover, we study the influence of the coupling strength between the two giant atoms on the locations of the reflection peaks and the transmission window in the single-photon scattering spectra. Finally, we consider the single-photon scattering when the system is in the non-Markovian regime, where the propagating time of the single photon between two neighboring coupling points is nonnegligible. In this case, the scattering spectra are characterized by more complicated line shapes due to the joint influence of the strong detuning-dependent and constant-part phases.

The rest of this paper is organized as follows. In Sec.~\ref{Physical model}, we present the Hamiltonians and the equations of motion for probability amplitudes. In Secs.~\ref{SPTAGM} and ~\ref{NonMarkovian}, we study the single-photon scattering in the Markovian and non-Markovian regimes, respectively. In particular, we consider three different configurations of couplings between the giant molecule and the waveguide. Finally, we conclude this paper in Sec.~\ref{Conclusions}.
\begin{figure}[tbp]
\center\includegraphics[width=0.5\textwidth]{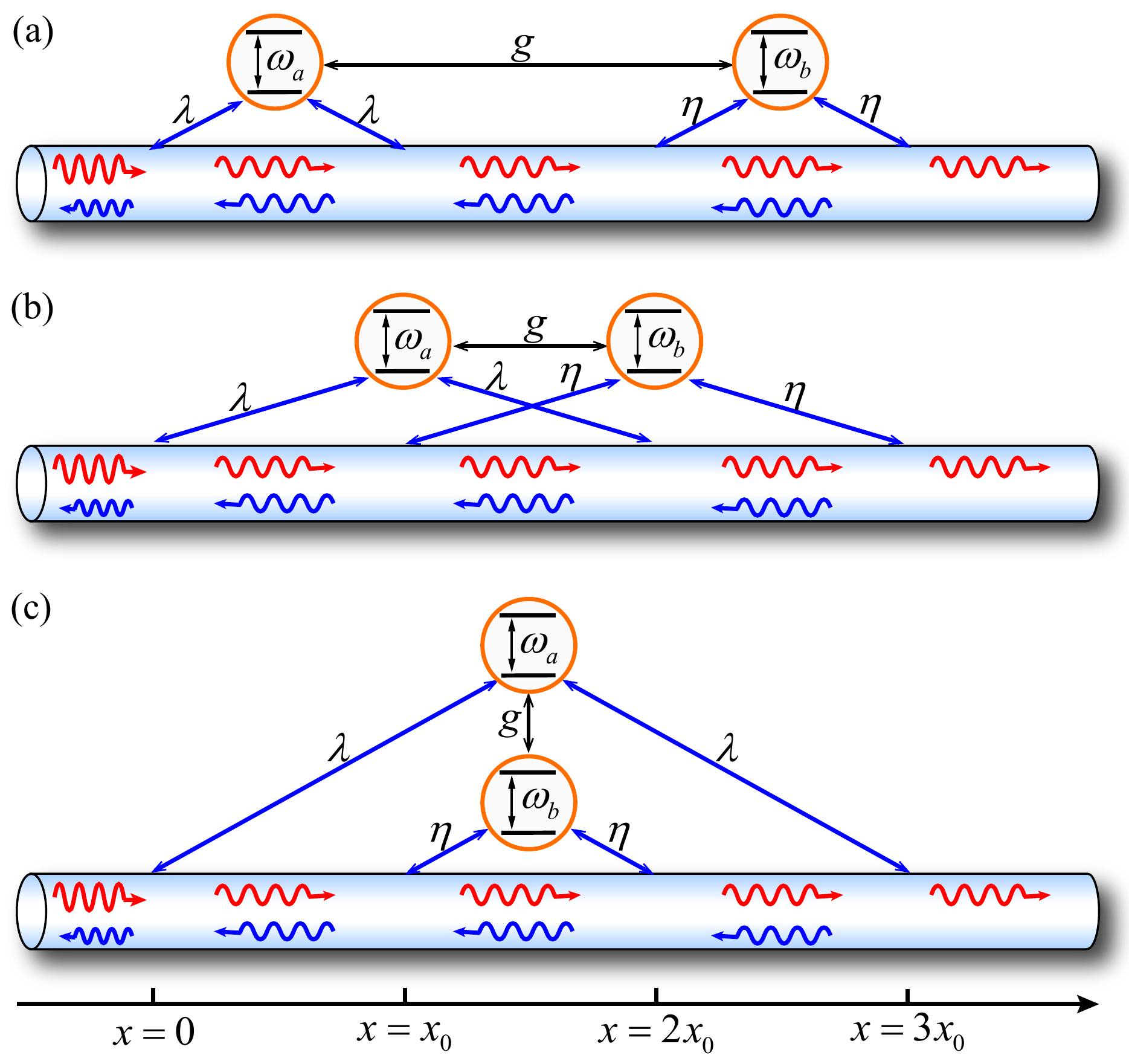}
\caption{Schematic of the giant-molecule waveguide-QED system. The giant molecule consists of two coupled giant atoms $a$ and $b$. The coupling between the giant atoms and the waveguide has three different coupling configurations:  (a) separated-coupling case,  (b) braided-coupling case,  and (c) nested-coupling case.}
\label{model_v1}
\end{figure}

\section{Hamiltonians and equations of motion}\label{Physical model}
We consider a giant-molecule waveguide-QED system, in which a giant artificial molecule couples to a one-dimensional waveguide. Here, the giant molecule consists of two coupled giant atoms $a$ and $b$. Each giant atom is coupled to the waveguide via two coupling points. For simplicity, hereafter we assume that the distances between any two adjacent coupling points are equal. According to the different arrangements of the coupling points between the giant atoms and the waveguide, we can construct three different coupling configurations~\cite{Kockum2018} of the coupled giant-molecule-waveguide system, including the separated-coupling case [Fig.~\ref{model_v1}(a)], the braided-coupling case [Fig.~\ref{model_v1}(b)], and the nested-coupling case [Fig.~\ref{model_v1}(c)]. In the rotating-wave approximation, the Hamiltonian of the total system reads  ($\hbar=1$)
\begin{eqnarray}
\hat{H} &=&\hat{H}_{s}+\hat{H}_{w}+\hat{V}_{i},\notag \\
\hat{H}_{s} &=&\omega _{a}\hat{\sigma}_{a}^{+}\hat{\sigma}_{a}^{-}+\omega _{b}\hat{\sigma}_{b}^{+}\hat{\sigma}_{b}^{-}+g( \hat{\sigma}_{a}^{-}\hat{\sigma}_{b}^{+}+\hat{\sigma}_{b}^{-}\hat{\sigma}_{a}^{+}), \notag \\
\hat{H}_{w} &=&i\upsilon _{g}\int dx\left( \hat{c}_{L}^{\dagger }\left( x\right)\frac{\partial }{\partial x}\hat{c}_{L}\left( x\right) -\hat{c}_{R}^{\dagger
}\left( x\right) \frac{\partial }{\partial x}\hat{c}_{R}\left( x\right)\right), \notag \\
\hat{V}_{i}&=&\lambda \int dxP_{i}\left( x\right) \left[\hat{\sigma}_{a}^{+}\hat{c}_{R}\left( x\right) +\hat{\sigma}_{a}^{+}\hat{c}_{L}\left( x\right) +\text{H.c.}\right] \notag \\
&&+\eta \int dxQ_{i}\left( x\right) \left[\hat{\sigma}_{b}^{+}\hat{c}_{R}\left(x\right) +\hat{\sigma}_{b}^{+}\hat{c}_{L}\left( x\right) +\text{H.c.}\right],
\end{eqnarray}
where $P_{i}(x)=\delta(x) +\delta(x-m_{i}x_{0})$ and $Q_{i}(x)=\delta(x-n_{i}x_{0}) +\delta (x-l_{i}x_{0})$ with $i=S,B$, and $N$ corresponding to the three coupling configurations  in Figs.~\ref{model_v1}(a),~\ref{model_v1}(b), and~\ref{model_v1}(c), respectively. The $\delta(x)$ is the Dirac delta function. The Hamiltonian $\hat{H}_{s}$ includes the free and interaction parts of the two giant atoms $a$ and $b$, where $\omega_{a}$ ($\omega_{b}$) denotes the transition frequency of the two-level atom  $a$ ($b$), $g$ represents the coupling strength between the atoms $a$ and $b$, and $\hat{\sigma}_{a}^{+}$ ($\hat{\sigma}_{b}^{+}$) = $(\hat{\sigma}_{a}^{-})^{\dagger
}[(\hat{\sigma}_{b}^{-})^{\dagger }]$ = $|e\rangle
_{aa}\langle g|$ ($|e\rangle_{bb}\langle g|$) is the raising operator of the atom $a$ ($b$) with the excited state $|e\rangle_{a}$ ($|e\rangle_{b}$) and the ground state $|g\rangle_{a}$ ($|g\rangle_{b}$). The term $\hat{H}_{w}$ is the free Hamiltonian of the waveguide with the  group velocity $\upsilon_{g}$.  The operators $\hat{c}_{R}^{\dagger}(x)$ $[\hat{c}_{R}(x)]$ and $\hat{c}_{L}^{\dagger}(x)$ $[\hat{c}_{L}(x)]$ are the field operators describing the creation (annihilation) of a right- and left-propagating photon at position $x$ in the waveguide, respectively.  The term $\hat{V}_{i}$ describes the interactions between the waveguide and the giant atoms, where $\lambda$ ($\eta$) is the coupling strength between the waveguide and the giant atom $a$ ($b$). The coefficients $m_{i}$, $n_{i}$, and $l_{i}$ in $P_{i}(x)$ and $Q_{i}(x)$ for these three different coupling configurations are given by
\begin{subequations}
\begin{align}
m_{S}=&1,\hspace{0.5cm}n_{S}=2,\hspace{0.5cm}l_{S}=3, \label{Conts for Separated case} \\
m_{B}=&2,\hspace{0.5cm}n_{B}=1,\hspace{0.5cm}l_{B}=3,  \label{Conts for Bradided case} \\
m_{N} =&3,\hspace{0.5cm}n_{N}=1,\hspace{0.5cm}l_{N}=2.\label{Conts for Nested case}
\end{align}
\end{subequations}

We assume that a single photon with energy $E$ is injected from the left-hand side of the waveguide. Since the total excitation number of the system is a conserved quantity, in the single-excitation subspace the eigenstate of the system can be expressed as
\begin{eqnarray}
\label{wavefunction of systm}
\left\vert \psi \right\rangle  &=&\int dx[\phi_{R}(x)\hat{c}_{R}^{\dagger}(x)+\phi_{L}(x) \hat{c}
_{L}^{\dagger}(x) ] \left\vert \emptyset \right\rangle\notag \\
&&+u_{a}\hat{\sigma}_{a}^{+}\left\vert \emptyset \right\rangle +u_{b}\hat{\sigma}_{b}^{+}\left\vert \emptyset \right\rangle,
\end{eqnarray}
where $\phi_{R}(x)$ [$\phi_{L}(x)$] is the single-photon wave function of the right-propagating (left-propagating) mode at position $x$, $u_{a}$ ($u_{b}$) is the probability amplitude of the giant atom $a$ ($b$), and $\left\vert \emptyset \right\rangle$ represents the vacuum state, which means that there are no photons in the waveguide and the two atoms are in their ground states. By substituting Eq.~(\ref{wavefunction of systm}) into the stationary Schr\"{o}dinger equation $\hat{H}\left\vert \psi \right\rangle =E\left\vert \psi \right\rangle$, we can obtain the following equations of motion for the probability amplitudes
\begin{eqnarray}
\label{motion of Eq for amplitudes}
-i\upsilon_{g}\frac{\partial \phi _{R}\left( x\right) }{\partial x}+\lambda
u_{a}P_{i}(x)+\eta u_{b}Q_{i}(x) &=&E\phi _{R}\left( x\right), \notag \\
i\upsilon_{g}\frac{\partial \phi _{L}\left( x\right) }{\partial x}+\lambda
u_{a}P_{i}(x)+\eta u_{b}Q_{i}(x) &=&E\phi _{L}\left( x\right), \notag \\
\omega _{a}u_{a}+gu_{b}+\lambda A_{i} &=&Eu_{a},\notag \\
\omega _{b}u_{b}+gu_{a}+\eta B_{i} &=&Eu_{b},
\end{eqnarray}
where we introduce
\begin{eqnarray}
A_{i} &=&\phi _{R}\left( 0\right) +\phi _{L}\left( 0\right) +\phi _{R}\left(
m_{i}x_{0}\right) +\phi _{L}\left( m_{i}x_{0}\right),\notag \\
B_{i} &=&\phi _{R}\left( n_{i}x_{0}\right) +\phi _{L}\left(n_{i}x_{0}\right) +\phi _{R}\left( l_{i}x_{0}\right) +\phi _{L}\left(l_{i}x_{0}\right).
\end{eqnarray}
\begin{figure*}[tbp]
\center\includegraphics[width=0.99\textwidth]{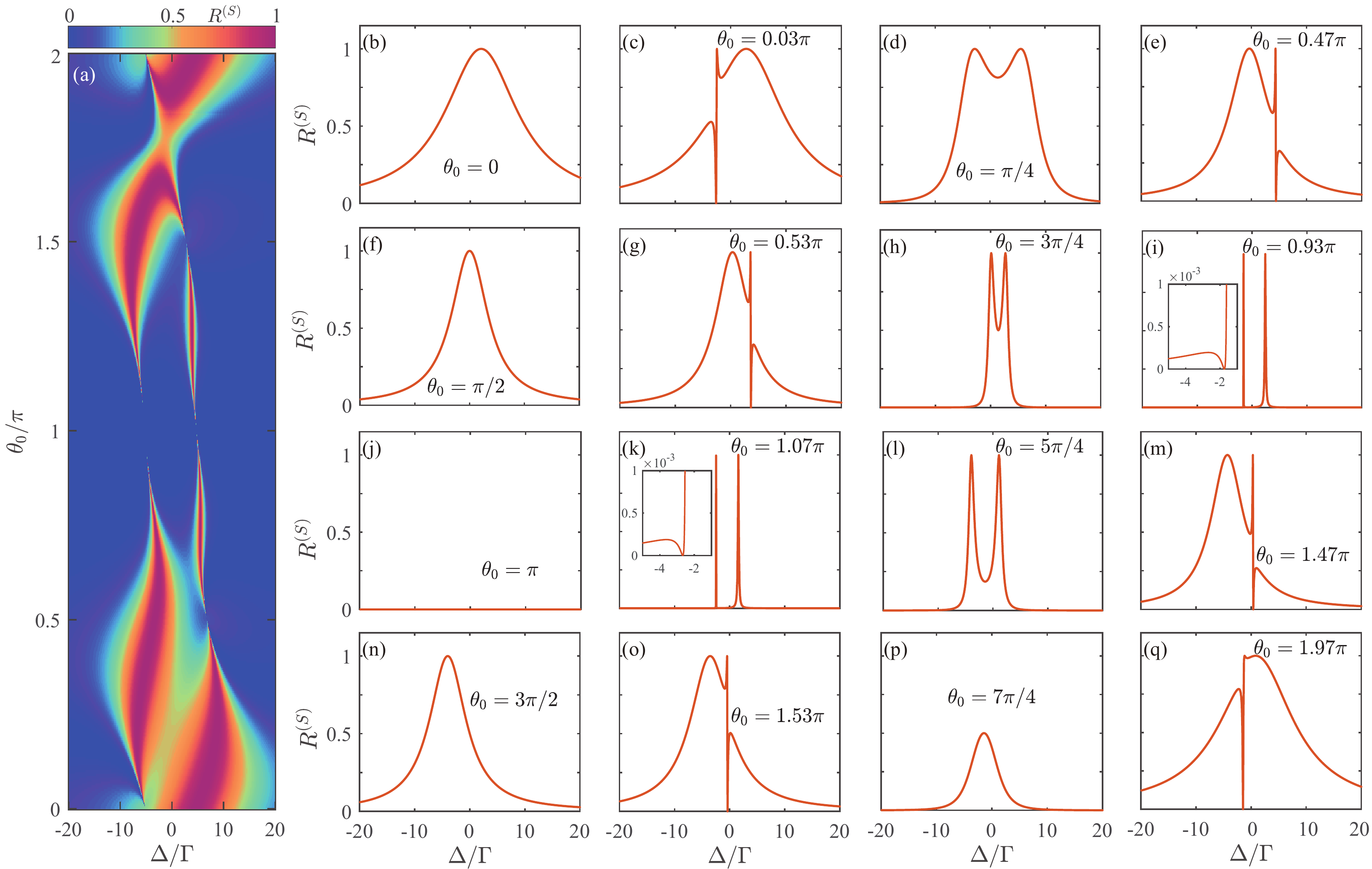}
\caption{(a) Reflection coefficient $R^{(S)}$ as a function of the detuning $\Delta$ and the phase shift $\theta_{0}$. The profiles of panel (a) are shown by the curves in (b)-(q) at different phases: (b) $\theta_{0}=0$, (c) $\theta_{0}=0.03\pi$, (d) $\theta_{0}=\pi/4$, (e) $\theta_{0}=0.47\pi$, (f) $\theta_{0}=\pi/2$, (g) $\theta_{0}=0.53\pi$, (h) $\theta_{0}=3\pi/4$, (i) $\theta_{0}=0.93\pi$, (j) $\theta_{0}=\pi$, (k) $\theta_{0}=1.07\pi$, (l) $\theta_{0}=5\pi/4$, (m) $\theta_{0}=1.47\pi$, (n) $\theta_{0}=3\pi/2$, (o) $\theta_{0}=1.53\pi$, (p) $\theta_{0}=7\pi/4$, and (q) $\theta_{0}=1.97\pi$. In all panels, the coupling strength is $g/\Gamma=2$.}
\label{RSvsDelta}
\end{figure*}

In this paper, we consider the single-photon scattering in the waveguide. Assuming that initially a photon with wave vector $k$ ($k>0$) is injected from the left-hand side of the waveguide and the two atoms are in their ground states $|g\rangle_{a}$ and $|g\rangle_{b}$, then the wave functions in Eq.~(\ref{wavefunction of systm}) take the form
\begin{subequations}
\begin{align}
\phi _{R}(x)=&e^{ikx}[\Theta (-x)+t_{1}^{(f)}\Theta (x)\Theta
(x_{0}-x)\nonumber \\
&+t_{2}^{(f)}\Theta (x-x_{0})\Theta (2x_{0}-x)  \nonumber \\
&+t_{3}^{(f)}\Theta (x-2x_{0})\Theta (3x_{0}-x)+t^{(f)}\Theta(x-3x_{0})],  \label{amplitude expressions R}\\
\phi _{L}(x)=&e^{-ikx}[r^{(f)}\Theta (-x)+r_{1}^{(f)}\Theta (x)\Theta
(x_{0}-x)  \nonumber \\
&+r_{2}^{(f)}\Theta (x-x_{0})\Theta (2x_{0}-x) \nonumber \\
&+r_{3}^{(f)}\Theta (x-2x_{0})\Theta (3x_{0}-x)],\label{amplitude expressions L}
\end{align}
\end{subequations}
with $f=S,B$, and $N$. In Eqs.~(\ref{amplitude expressions R}) and~(\ref{amplitude expressions L}), $t^{(f)}$ ($r^{(f)}$) represents the transmission (reflection) amplitude for the last (first) coupling points, $t^{(f)}_{j=1,2,3}$ ($r^{(f)}_{j=1,2,3}$) is the transmission (reflection) amplitude for the $j$th [$(j+1)$th] coupling point, and $\Theta(x)$ is the Heaviside step function with $\Theta(0)=1/2$. Note that the three different coupling configurations of the artificial-giant-molecule waveguide-QED system in Fig.~\ref{model_v1} have the equivalent form as Eqs.~(\ref{amplitude expressions R}) and~(\ref{amplitude expressions L})  due to the same number of equidistant coupling points  in the waveguide.

\section{Single-photon scattering in the waveguide coupled to the giant molecule with three configurations}\label{SPTAGM}
In this section, we study single-photon scattering in the giant-artificial-molecule waveguide-QED system. In particular, the couplings between the giant molecule and the waveguide have three different coupling configurations. Concretely, we calculate the corresponding single-photon transmission and reflection amplitudes and then obtain the transmission and reflection coefficients.
\subsection{Separated-coupling case}
To start with, we consider the separated-coupling case, as described in Fig.~\ref{model_v1}(a). In this case, the coefficients $m_{i}$, $n_{i}$, and $l_{i}$ in $P_{i}(x)$ and $Q_{i}(x)$ are given by Eq.~(\ref{Conts for Separated case}). Substituting these coefficients and Eqs.~(\ref{amplitude expressions R}) and~(\ref{amplitude expressions L}) into Eq.~(\ref{motion of Eq for amplitudes}),  one can obtain the following equations
\begin{eqnarray}
\label{t and r eq1}
i\upsilon_{g}\left(1-t_{1}^{(S)}\right)+\lambda u_{a}&=&0,  \notag \\
i\upsilon_{g}\left(r_{1}^{(S)}-r^{(S)}\right)+\lambda u_{a}&=&0,  \notag \\
i\upsilon_{g}\left(t_{1}^{(S)}-t_{2}^{(S)}\right)e^{ikx_{0}}+\lambda u_{a}&=&0,  \notag \\
i\upsilon_{g}\left(r_{2}^{(S)}-r_{1}^{(S)}\right)e^{-ikx_{0}}+\lambda u_{a}&=&0,  \notag \\
i\upsilon_{g}\left(t_{2}^{(S)}-t_{3}^{(S)}\right)e^{2ikx_{0}}+\eta u_{b}&=&0,  \notag \\
i\upsilon_{g}\left(r_{3}^{(S)}-r_{2}^{(S)}\right)e^{-2ikx_{0}}+\eta u_{b}&=&0,  \notag \\
i\upsilon_{g}\left(t_{3}^{(S)}-t^{(S)}\right)e^{3ikx_{0}}+\eta u_{b}&=&0 ,  \notag \\
-i\upsilon_{g}r_{3}^{(S)}e^{-3ikx_{0}}+\eta u_{b}&=&0 ,
\end{eqnarray}
and
\begin{eqnarray}
\label{t and r eq2}
\Delta u_{a}&=&gu_{b}+\frac{\lambda }{2}\left[ 1+t_{1}^{(S)}+
r^{(S)}+r_{1}^{(S)}+\left(t_{1}^{(S)}+t_{2}^{(S)}\right)e^{ikx_{0}}
\right.   \notag \\
&&\left. +\left(r_{1}^{(S)}+r_{2}^{(S)}\right)e^{-ikx_{0}}\right] ,  \notag \\
\Delta u_{b} &=&gu_{a}+\frac{\eta }{2}\left[\left(t_{2}^{(S)}+t_{3}^{(S)}\right)e^{2ikx_{0}}+\left(r_{2}^{(S)}+r_{3}^{(S)}\right)e^{-2ikx_{0}}
\right.   \notag \\
&&\left. +\left(t_{3}^{(S)}+t\right)e^{3ikx_{0}}+r_{3}^{(S)}e^{-3ikx_{0}}\right] ,
\end{eqnarray}
where  $\Delta =E-\omega_{a}=E-\omega_{b}=E-\Omega$ is the detuning between the resonance frequency of the photon propagating in the waveguide and the transition frequency associated with $|e\rangle_{a}\leftrightarrow|g\rangle_{b}$ $(|e\rangle_{b}\leftrightarrow|g\rangle_{b})$ of atom $a$ ($b$). Here, we assume that $\omega_{a}=\omega_{b}=\Omega$. In this situation, the two giant atoms are resonant with the traveling photons with wave vector $k$. According to Eqs.~(\ref{t and r eq1}) and~(\ref{t and r eq2}), the transmission and reflection amplitudes in this case can be obtained as
\begin{widetext}
\begin{subequations}
\begin{align}
t^{(S)}&=\frac{-\left[ \Delta -2\Gamma _{1}\sin \theta \right] \left[ \Delta-2\Gamma _{2}\sin \theta \right] +g^{2}+4g\sqrt{\Gamma _{1}\Gamma _{2}}\sin
\left( 2\theta \right) \left( 1+\cos \theta \right) }{\left[ i\Delta -\left(\Gamma _{1}+\Gamma _{2}\right) \left( 1+e^{i\theta }\right) \right]^{2}
-\left( \Gamma _{1}-\Gamma _{2}\right) ^{2}\left( 1+e^{i\theta }\right)^{2}
-\left[ig+\sqrt{\Gamma _{1}\Gamma _{2}}e^{i\theta }\left( 1+e^{i\theta}\right) ^{2}\right] ^{2}},\label{ts}\\
r^{(S)}&=\frac{4ie^{3i\theta }\cos \left( \frac{\theta }{2}\right) ^{2}\left \{
\Delta \left[ \left( \Gamma _{1}+\Gamma _{2}\right) \cos \left( 2\theta
\right) +i\left( \Gamma _{2}-\Gamma _{1}\right) \sin \left( 2\theta \right)
\right] +4\Gamma _{1}\Gamma _{2}\left[ \sin \left( \theta \right) +\sin
\left( 2\theta \right) \right] +2g\sqrt{\Gamma _{1}\Gamma _{2}}\right \} }{
\left[ i\Delta -\left( \Gamma _{1}+\Gamma _{2}\right) \left( 1+e^{i\theta
}\right) \right] ^{2}-\left( \Gamma _{1}-\Gamma _{2}\right) ^{2}\left(
1+e^{i\theta }\right) ^{2}-\left[ ig+\sqrt{\Gamma _{1}\Gamma _{2}}e^{i\theta
}\left( 1+e^{i\theta }\right) ^{2}\right] ^{2}},\label{rs}
\end{align}
\end{subequations}
\end{widetext}
where we introduce the decay rate $\Gamma_{1}=\lambda^{2}/\upsilon_{g}$ ($\Gamma_{2}=\eta^{2}/\upsilon_{g}$) related to the decay from the excited state $|e\rangle_{a}$ ($|e\rangle_{b}$) to the ground state $|g\rangle_{a}$ ($|g\rangle_{b}$). We point out that hereafter we neglect the intrinsic dissipation rate $\kappa$ of the giant atoms decaying out of the waveguide modes. This approximation is valid because the intrinsic dissipation rate $\kappa$ is much smaller than the decay rates $\Gamma_{1}$ and $\Gamma_{2}$ of the giant atoms in typical experiments~\cite{Kannan2020}. Therefore, we do not include the dissipation rate $\kappa$ of the giant atoms in the following discussions. Note that the effect of the dissipation rate $\kappa$ can also be estimated by replacing $\Delta$ with $\Delta+i\kappa$ in the transmission and reflection amplitudes.

In Eqs.~(\ref{ts}) and~(\ref{rs}), we introduce $\theta=kx_{0}$ to denote the accumulated phase shift of the photon with wave vector $k$ propagating between two adjacent coupling points in the waveguide. Based on the relations $\Delta=E-\Omega$ and $E=\upsilon_{g}k$, the phase shift $\theta$ can be written as
\begin{equation}
\theta=\tau\Delta+\theta_{0},\label{theta}
\end{equation}
where $\tau=x_{0}/\upsilon_{g}$ is defined as the propagating time of the single photon between two neighboring coupling points and $\theta_{0}=x_{0}\Omega/\upsilon_{g}$ is a constant part. When the propagating time $\tau$ is comparable to or larger than the atomic lifetime $1/\Gamma$, the retarded effect induced by $\tau$ cannot be neglected. Therefore, the giant atoms enter the so-called non-Markovian regime.

In Sec.~\ref{SPTAGM}, we will focus on single-photon scattering in the Markovian regime, in which the propagating time satisfies the condition $\tau\Gamma\ll 1$~\cite{Guo2017}, and hence the detuning-dependent terms $\tau\Delta$ in Eq.~(\ref{theta}) can be neglected. In this case, the phase shift $\theta$ in Eqs.~(\ref{ts}) and~(\ref{rs}) can be replaced with $\theta_{0}$. In principle, the phase shift $\theta_{0}=x_{0}\Omega/\upsilon_{g}$  can be adjusted in the region of $\theta_{0}\in[0,2\pi]$ by changing the distant between adjacent coupling points and the frequency of the giant atom. In the following discussions, we focus on the reflection coefficient $R^{(S)}=|r^{(S)}|^{2}$, since the reflection coefficient $R^{(S)}$ and the transmission coefficient $T^{(S)}=|t^{(S)}|^{2}$ satisfy the relation $R^{(S)}+T^{(S)}=1$. Note that the single-photon scattering in the non-Markovian regime will be considered in Sec.~\ref{NonMarkovian}.

For simplicity, hereafter we consider the case where the two giant atoms have the same decay rates, i.e., $\Gamma_{1}=\Gamma_{2}=\Gamma$, then the amplitudes given in Eqs.~(\ref{ts}) and~(\ref{rs}) can be simplified as
\begin{widetext}
\begin{subequations}
\begin{align}
t^{(S)}&=\frac{- \left[ \Delta -2\Gamma \sin \theta_{0}\right]^{2}+g^{2}+4g\Gamma\sin(2\theta_{0})(1+\cos\theta_{0}) }{\left[ i\Delta -2\Gamma \left( 1+e^{i\theta_{0}}\right) \right] ^{2}-
\left[ ig+\Gamma e^{i\theta_{0}}(1+e^{i\theta_{0}}) ^{2}\right] ^{2}},
\label{tsGammaeq} \\
r^{(S)}&=\frac{4i\Gamma e^{3i\theta_{0}}\cos\left(\frac{\theta_{0}}{2}\right) ^{2}\left \{ 2\Delta\cos(2\theta_{0})+4\Gamma
\left[ \sin\theta_{0}+\sin(2\theta_{0}) \right]+2g\right \}}{\left[ i\Delta -2\Gamma(1+e^{i\theta_{0}})\right] ^{2}-\left[ ig+\Gamma e^{i\theta_{0}}(1+e^{i\theta_{0}}) ^{2}\right]^{2}}.\label{rsGammaeq}
\end{align}
\end{subequations}
\end{widetext}
We point out that when the coupling between the two coupled atoms is turn off, i.e., $g=0$, Eqs.~(\ref{tsGammaeq}) and~(\ref{rsGammaeq}) can be reduced to the results of the two separated giant atoms described in Ref.~\cite{Jia2021}.

In Fig.~\ref{RSvsDelta}(a), we show the reflection coefficient $R^{(S)}$ as a function of the detuning $\Delta$ and the phase shift $\theta_{0}$.  One can see that the width and the location of the reflection peak can be modified by adjusting the phase shift $\theta_{0}$, and that the scattering characteristics is phase dependent with a period of $2\pi$. Different from the results in Ref.~\cite{Jia2021}, the coupling strength between the two giant atoms leads to the relation $R^{(S)}(\Delta,\theta_{0})\neq R^{(S)}(-\Delta,2\pi-\theta_{0})$. Therefore, the phase shift considered here is $\theta_{0}\in[0,2\pi]$. As $\theta_{0}$ goes from 0 to $2\pi$, the width of the reflection peak exhibits nonmonotonic changes. According to Eqs.~(\ref{tsGammaeq}) and~(\ref{rsGammaeq}), we find that the two reflection peaks with $R^{(S)}=1$ (the complete reflection) are located at $\Delta=2\Gamma\sin\theta_{0}\pm\sqrt{g^{2}+4g\Gamma\sin(2\theta_{0})(1+\cos\theta_{0})}$ under the condition $g^{2}+4g\Gamma\sin(2\theta_{0})(1+\cos\theta_{0})>0$. The reflection spectrum takes the minimal value $R^{(S)}=0$ at $\Delta=-\left\{2\Gamma\left[\sin(\theta_{0})+\sin(2\theta_{0})\right]+g\right\}/\cos(2\theta_{0})$. In addition, the two peaks of the reflection coefficient always appear at the same side of the minimal value [namely the right-hand side when $\theta_{0}\rightarrow 0$ and $2\pi$, and the left-hand side when $\theta_{0}\rightarrow\pi/2$ and $3\pi/2$].

In Figs.~\ref{RSvsDelta}(b)-\ref{RSvsDelta}(q) we plot the profiles of Fig.~\ref{RSvsDelta}(a) at different phase shifts in the region of $\theta_{0}\in [0,2\pi]$. We find that, when $\theta_{0}=0$, $\pi/2$, and $3\pi/2$ [see Figs.~\ref{RSvsDelta}(b),~\ref{RSvsDelta}(f), and~\ref{RSvsDelta}(n)], the reflection spectra are characterized by the Lorentzian line shapes. To see this feature more clearly, we substitute $\theta_{0}=0$, $\pi/2$, and $3\pi/2$ into Eq.~(\ref{rsGammaeq}) and obtain $r^{(S)}=-8i\Gamma /\left( \Delta -g+8i\Gamma\right)$, $4\Gamma /\left( \Delta +g-2\Gamma +4i\Gamma \right)$,  and $-4\Gamma /\left( \Delta +g+2\Gamma +4i\Gamma \right)$, respectively. The corresponding reflection peaks appear at $\Delta=g$, $2\Gamma-g$, and $-2\Gamma-g$ with the widths $16\Gamma$, $8\Gamma$, and $8\Gamma$, respectively.

When $\theta_{0}\in(0,\pi/2)\cup(\pi/2,\pi)\cup(\pi,3\pi/2)$, $\theta_{0}=1.53\pi$, and $\theta_{0}=1.97\pi$, there are two peaks of the reflection spectra as shown in Figs.~\ref{RSvsDelta}(c)-\ref{RSvsDelta}(e) ,~\ref{RSvsDelta}(g)-\ref{RSvsDelta}(i),~\ref{RSvsDelta}(k)-\ref{RSvsDelta}(m), ~\ref{RSvsDelta}(o), and~\ref{RSvsDelta}(q), which indicate that the locations and widths of the peaks depend strongly on the phase $\theta_{0}$. From Figs.~\ref{RSvsDelta}(d),~\ref{RSvsDelta}(h),~\ref{RSvsDelta}(l), and~\ref{RSvsDelta}(p),  we see that the reflection spectra are symmetrical to $\Delta=2\Gamma\sin\theta_{0}$ when $\theta_{0}=\pi/4$, $3\pi/4$, $5\pi/4$, and $7\pi/4$. According to the condition for the existence of two complete reflection peaks, we can obtain two roots for $g^{2}+4g\Gamma\sin(2\theta_{0})(1+\cos\theta_{0})=0$ at $g/\Gamma=2$, which are $\theta_{0}\approx1.57\pi$ and $1.95\pi$. When $\theta_{0}\in(1.57\pi,1.95\pi)$, the injected single photon can only be reflected partially. In particular, when  $\theta_{0}=7\pi/4$ and $g/\Gamma=2$, the reflection coefficient $R^{S}=0.5$ [see Fig.~\ref{RSvsDelta}(p)], which means that this giant-molecule emitter can act as a 50:50 beam splitter for a single-photon by properly choosing the coupling strength $g$ and the phase shift $\theta_{0}$. In some regions, we also find that the reflection spectra near their minimal values can exhibit the phenomenon of Fano resonance, where the minimal values are determined by $\Delta=-\left\{2\Gamma\left[\sin(\theta_{0})+\sin(2\theta_{0})\right]+g\right\}/\cos(2\theta_{0})$. Therefore, the position of the Fano minimum can be regarded as an estimation of the coupling strength between the two atoms. As shown in Figs.~\ref{RSvsDelta}(c), ~\ref{RSvsDelta}(e),~\ref{RSvsDelta}(g),~\ref{RSvsDelta}(i),~\ref{RSvsDelta}(k), ~\ref{RSvsDelta}(m),~\ref{RSvsDelta}(o), and~\ref{RSvsDelta}(q), the reflection spectra are similar to the asymmetric Fano line shapes when $\theta_{0}$ takes some specific values. Meanwhile, the reflection peaks always appear at the same side of the dip. One can prove that the reflection amplitude in Eq.~(\ref{rsGammaeq}) can be decomposed as the superposition of two Lorentzian spectra with shifted centers of symmetry, that is $r^{(S)}=r_{+}^{(S)}+r_{-}^{(S)}$ with
\begin{equation}
\label{two Lorentz}
r_{\pm }^{(S)}=\frac{\pm e^{3i\theta_{0}}\Gamma _{\pm }}{i\left(\tilde{\Delta}_{\mp}-\Lambda _{\pm }\right) -\Gamma _{\pm }}.
\end{equation}
Here the effective detuning, the peak points, and the half widths are defined, respectively, as
\begin{subequations}
\begin{align}
\tilde{\Delta}_{\mp } &=\Delta \mp g , \\
\Lambda_{\pm }&=2\Gamma\sin\theta_{0}\left(1\pm 2\cos\theta_{0}\pm 2\cos^{2}\theta_{0}\right), \\
\Gamma _{\pm }& =2\Gamma \left(1+\cos\theta_{0}\right) \left[ 1\pm \cos(2\theta_{0})\right] .
\end{align}
\end{subequations}
When the half widths satisfy the condition $\Gamma_{+}\gg\Gamma_{-}$ (or $\Gamma_{-}\gg\Gamma_{+}$), we can obtain $\Lambda_{+}\approx\Lambda_{-}$ for the peak points. Therefore the reflection coefficient near $\Lambda_{\pm}$ can be approximated as a standard Fano line shape, that is,
\begin{equation}
\label{Fano lineshape}
R^{(S)}\approx \mathcal{C_{\pm}}\frac{(q_{\pm}+\epsilon_{\pm}) ^{2}}{1+\epsilon^{2}_{\pm}},
\end{equation}
where we introduce the reduced detunings $\epsilon_{\pm}=(\tilde{\Delta}_{\pm }-\Lambda_{\mp})/\Gamma_{\mp}$, the asymmetric parameters $q_{\pm}=(\Lambda_{\pm }-\Lambda_{\mp }) /\Gamma_{\pm}$, and the modified coefficients $\mathcal{C_{\pm}}=\Gamma_{\pm}^{2}/[(\Lambda_{\pm}-\Lambda_{\mp})^{2}+\Gamma_{\pm}^{2}]$. Here, the upper and lower indices in these subscripts $\pm$ correspond to the cases of $\Gamma_{+}\gg\Gamma_{-}$ and $\Gamma_{-}\gg\Gamma_{+}$, respectively. On the one hand, when $\Gamma_{+}>15\Gamma_{-}$, we have $\Gamma_{+}\gg\Gamma_{-}$. Then the reflection spectra can be approximated as the Fano line shapes at $\theta_{0}\in(0,0.08\pi)\cup(0.92\pi,\pi)\cup(\pi,1.08\pi)\cup(1.92\pi,\pi)$ [see Figs.~\ref{RSvsDelta}(c),~\ref{RSvsDelta}(i),~\ref{RSvsDelta}(k), and~\ref{RSvsDelta}(q)]. On the other hand, for $\Gamma_{-}>15\Gamma_{+}$, we obtain the Fano line shapes when $\theta_{0}\in(0.42\pi,0.5\pi)\cup(0.5\pi,0.58\pi)\cup(1.42\pi,1.5\pi)\cup(1.5\pi,1.58\pi)$ [see Figs.~\ref{RSvsDelta}(e),~\ref{RSvsDelta}(g),~\ref{RSvsDelta}(m), and~\ref{RSvsDelta}(o)].
\begin{figure}[tbp]
\center\includegraphics[width=0.48\textwidth]{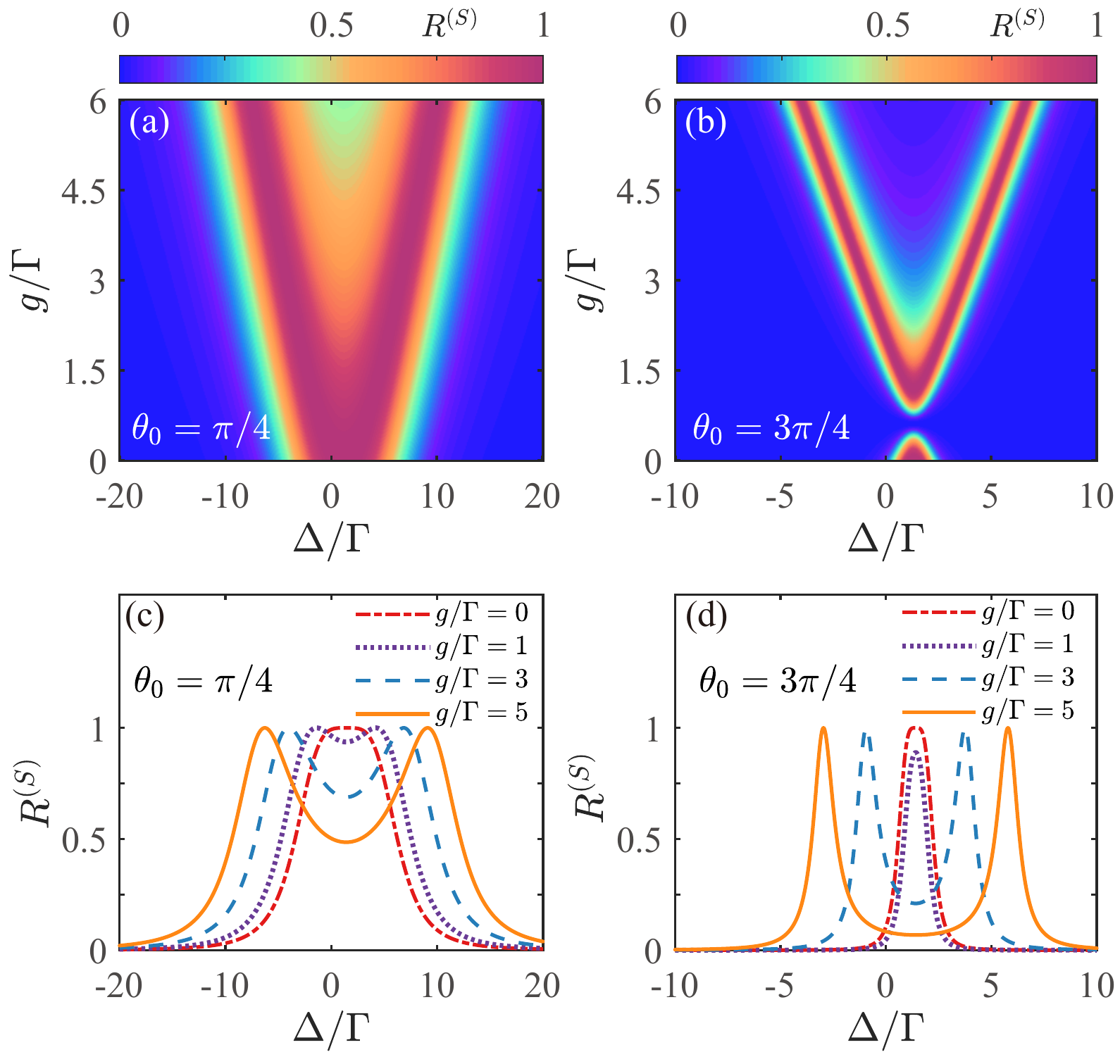}
\caption{(a) and (b) Reflection coefficient $R^{(S)}$ vs $\Delta$ and $g$ with $\theta_{0}=\pi/4$ and $3\pi/4$, respectively. The curves in (c) and (d) show the profiles of (a) and (b) at given values of $g$.}
\label{RSvsDeltaandg}
\end{figure}

To investigate the influence of the coupling between the two giant atoms on the single-photon scattering, we show the reflection coefficient $R^{(S)}$ as a function of $\Delta$ and $g$ with $\theta_{0}=\pi/4$ and $3\pi/4$ in Figs.~\ref{RSvsDeltaandg}(a) and~\ref{RSvsDeltaandg}(b), respectively. In principle, we can obtain the reflection spectra as functions of $\Delta$ and $g$ for all phase shifts $\theta_{0}$. In order to exhibit typical features of the reflection spectra, we take the phase shifts $\theta_{0}=\pi/4$ and $3\pi/4$ in the simulations. In these cases, the reflection spectra are symmetric to $\Delta=2\Gamma\sin\theta_{0}$ [see Figs.~\ref{RSvsDelta}(d) and~\ref{RSvsDelta}(h)]. Figure~\ref{RSvsDeltaandg}(a) shows that the number of the peaks in $R^{(S)}$ varies from 1 to 2 as $g$ increases from zero to $g/\Gamma=5$ when $\theta_{0}=\pi/4$, which can be seen more clearly from the profile of $R^{(S)}$ in Fig.~\ref{RSvsDeltaandg}(c). The separation of the two peaks and the width of the valleys of the spectra gradually increase with the increase of $g$. This means that the incident photon can transmit the waveguide with a certain probability by tuning the coupling strength between the two giant atoms. According to the locations of the two reflection peaks, we can obtain the separation between the two peaks, which is determined by $d^{(S)}=2\sqrt{g^{2}+4g\Gamma\sin(2\theta_{0})(1+\cos\theta_{0})}$. Therefore, in the case of $\theta_{0}=\pi/4$, it can be seen that the larger the coupling strength is, the wider the transmission is. However, this feature becomes different when $\theta_{0}=3\pi/4$, because $d^{(S)}$ is an imaginary number when $g/\Gamma\in(0,4-2\sqrt{2}]$, which leads to the reflection coefficient with only one peak located at $\Delta=2\Gamma\sin\theta_{0}$ [see the purple dotted line for $g/\Gamma=1$ in Fig.~\ref{RSvsDeltaandg}(d)]. In addition, when $\theta_{0}=3\pi/4$, the widths of the two reflection peaks are narrower than those in the case of $\theta_{0}=\pi/4$, as shown in Fig.~\ref{RSvsDeltaandg}(b). Similarly, we plot the profiles of $R^{(S)}$ in Fig.~\ref{RSvsDeltaandg}(d) when $\theta_{0}=3\pi/4$, which indicate that, in the region $g/\Gamma\in[0,4-2\sqrt{2}]$, there is only one reflection peak and the peak value first decreases and then increases to 1 as the coupling strength $g/\Gamma$ goes from 0 to $4-2\sqrt{2}$. As $g$ continues to increase, two peaks appear in the reflection coefficient and the separation between the peaks gradually enlarges.
\begin{figure*}[tbp]
\center\includegraphics[width=0.97\textwidth]{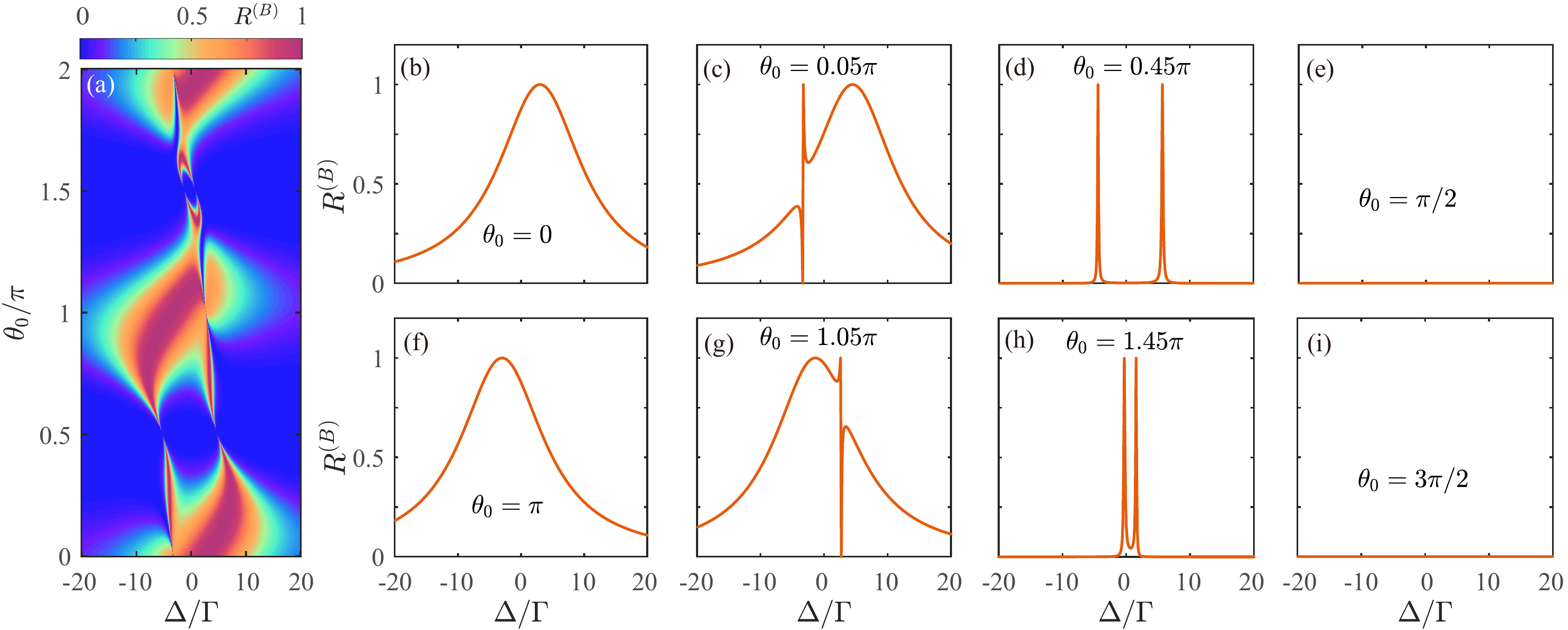}
\caption{(a) Reflection coefficient $R^{B}$ as a function of the detuning $\Delta$ and the phase shift $\theta_{0}$. The profiles of panel (a) are shown by the curves in (b)-(i) at different phases: (b) $\theta_{0}=0$, (c) $\theta_{0}=0.05\pi$, (d) $\theta_{0}=0.45\pi$, (e) $\theta_{0}=\pi/2$, (f) $\theta_{0}=\pi$, (g) $\theta_{0}=1.05\pi$, (h) $\theta_{0}=1.45\pi$, and (i) $\theta_{0}=3\pi/2$. In all panels, we choose $g/\Gamma=3$.}
\label{RBvsDelta}
\end{figure*}

\subsection{Braided-coupling case}
Next, we consider the braided-coupling case, as shown in Fig.~\ref{model_v1}(b).  In this case, the coefficients $m_{i}$, $n_{i}$, and $l_{i}$ in $P_{i}(x)$ and $Q_{i}(x)$ are given by Eq.~(\ref{Conts for Bradided case}). By substituting Eqs.~(\ref{amplitude expressions R}) and~(\ref{amplitude expressions L}) into Eq.~(\ref{motion of Eq for amplitudes}),  one can obtain the corresponding reflection and transmission amplitudes as
\begin{widetext}
\begin{subequations}
\begin{align}
t^{(B)}&=\frac{-\left[ \Delta -\left( \Gamma _{1}+\Gamma _{2}\right) \sin
\left( 2\theta \right) \right] ^{2}+\left( \Gamma _{1}+\Gamma _{2}\right)
^{2}\sin ^{2}\left( 2\theta \right) +4\Gamma _{1}\Gamma _{2}\sin ^{2}\theta
+g^{2}+2g\sqrt{\Gamma _{1}\Gamma _{2}}\left[ 3\sin \theta
+\sin \left( 3\theta \right) \right] }{\left[ i\Delta -\left( \Gamma _{1}
+\Gamma _{2}\right) \left( 1+e^{2i\theta}\right) \right]^{2}
-\left[ \left( \Gamma _{1}+\Gamma _{2}\right)^{2}
+\Gamma _{1}\Gamma _{2}e^{2i\theta }\right] \left(1+e^{2i\theta}\right) ^{2}
+4\Gamma _{1}\Gamma _{2}+g^{2}-2ig\sqrt{\Gamma _{1}\Gamma _{2}}\left( 3e^{i\theta }
+e^{3i\theta }\right)},\label{tb}\\
r^{(B)}&=\frac{4ie^{3i\theta }\cos ^{2}\theta \left[ \Delta \left( \Gamma
_{1}e^{-i\theta }+\Gamma _{2}e^{i\theta }\right) +4\Gamma _{1}\Gamma
_{2}\sin \theta +2g\sqrt{\Gamma _{1}\Gamma _{2}}\right] }{\left[ i\Delta
-\left( \Gamma _{1}+\Gamma _{2}\right) \left( 1+e^{2i\theta }\right) \right]
^{2}-\left[ \left( \Gamma _{1}+\Gamma _{2}\right) ^{2}+\Gamma _{1}\Gamma
_{2}e^{2i\theta }\right] \left( 1+e^{2i\theta }\right) ^{2}+4\Gamma
_{1}\Gamma _{2}+g^{2}-2ig\sqrt{\Gamma _{1}\Gamma _{2}}\left( 3e^{i\theta
}+e^{3i\theta }\right) }\label{rb}.
\end{align}
\end{subequations}
Considering the Markovian limit $\theta\approx\theta_{0}$ and the same decay rates $\Gamma_{1}=\Gamma_{2}=\Gamma$, Eqs.~(\ref{tb}) and~(\ref{rb}) are reduced to
\begin{subequations}
\begin{align}
t^{(B)}&=\frac{-\left[ \Delta -2\Gamma\sin(2\theta_{0}) \right]
^{2}+4\Gamma ^{2}\left[ \sin^{2}(2\theta_{0}) +\sin^{2}\theta_{0}\right] +g^{2}+2g\Gamma \left[3\sin\theta_{0}+\sin(3\theta_{0}) \right] }
{\left[ i\Delta -2\Gamma(1+e^{2i\theta_{0}}) \right] ^{2}-\left[ig+\Gamma (3e^{i\theta_{0}}+e^{3i\theta_{0}})\right] ^{2}},\label{tbGammaeq}\\
r^{(B)}&=\frac{8i\Gamma e^{3i\theta_{0}}\cos^{2}\theta_{0}\left[ \Delta\cos\theta_{0}
+2\Gamma\sin\theta_{0}+g\right] }{\left[ i\Delta -2\Gamma(1+e^{2i\theta_{0}}) \right] ^{2}-\left[ig+\Gamma(3e^{i\theta_{0}}+e^{3i\theta_{0}}) \right] ^{2}}.\label{rbGammaeq}
\end{align}
\end{subequations}
\end{widetext}
Figure~\ref{RBvsDelta}(a) shows the dependence of the reflection coefficient $R^{(B)}$ on $\Delta$ and $\theta_{0}$. Here we find that the period of the spectra is still $2\pi$, which is different from the case of $g=0$, in which the period of the spectra becomes $\pi$~\cite{Jia2021}.  According to Eqs.~(\ref{tbGammaeq}) and~(\ref{rbGammaeq}),  we know that the two reflection peaks with $R^{(B)}=1$ appear at $\Delta=2\Gamma \sin(2\theta_{0})\pm\sqrt{4\Gamma ^{2}\left[1-\cos\theta_{0}\cos(3\theta_{0})\right] +g^{2}+2g\Gamma \left[3\sin\theta_{0}+\sin(3\theta_{0}) \right]}$ in the scenario of $\theta_{0}\neq n\pi$ and under the condition $4\Gamma ^{2}\left[1-\cos\theta_{0}\cos(3\theta_{0})\right] +g^{2}+2g\Gamma \left[3\sin\theta_{0}+\sin(3\theta_{0}) \right]>0$. When $\theta_{0}=0$ and $(2n+1)\pi$ with integer $n$, the reflection amplitude given in Eq.~(\ref{rbGammaeq}) is reduced to $r^{(B)}=-8i\Gamma/\left(\Delta-g+8i\Gamma\right)$ and $-8i\Gamma /\left(\Delta+g+8i\Gamma\right)$, respectively,
which both behave as the Lorentzian line shapes with the same width $16\Gamma$ but different centers (at $\Delta=g$ and $-g$) [see Figs.~\ref{RBvsDelta}(b) and~\ref{RBvsDelta}(f)]. In terms of Eqs.~(\ref{tbGammaeq}) and~(\ref{rbGammaeq}), the minimum of the reflection coefficient occurs at $\Delta=-2\Gamma\tan\theta_{0}-g/\cos\theta_{0}$. In particular, one can find that the minimal value is always located either at the left-hand side [when $\theta_{0}\in(0,\pi/2)\cup(3\pi/2,2\pi)$] or the right-hand side [when $\theta_{0}\in(\pi/2,\pi)\cup(\pi,3\pi/2)$] of the two peaks of the spectra. To see the influence of the phase delay $\theta_{0}$ on the reflection coefficient $R^{(B)}$, we plot the profiles of Fig.~\ref{RBvsDelta}(a) in Figs.~\ref{RBvsDelta}(b)-\ref{RBvsDelta}(i) when $\theta_{0}$ takes some special values in $\theta_{0}\in(0,\pi/2)\cup(\pi,3\pi/2)$, due to $R^{(B)}(\Delta,\theta_{0})=R^{(B)}(-\Delta,\pi-\theta_{0})$ and $R^{(B)}(\Delta,\pi+\theta_{0})=R^{(B)}(-\Delta,2\pi-\theta_{0})$.

As shown in Figs.~\ref{RBvsDelta}(c) and~\ref{RBvsDelta}(g), the reflection spectra near its minimal value can also be characterized as the Fano line shape when the phase shift takes appropriate values. Similar to the separated-coupling case, the reflection amplitude in Eq.~(\ref{rbGammaeq}) can also be decomposed as $r^{(B)}=r^{(B)}_{+}+r^{(B)}_{-}$, where the Lorentzian spectra $r^{(B)}_{\pm}$ have the same form of Eq.~(\ref{two Lorentz}), with
\begin{subequations}
\begin{align}
\Lambda_{\pm }&=\Gamma\left[ 2\sin(2\theta_{0}) \pm 3\sin\theta_{0}\pm\sin(3\theta_{0})\right],  \\
\Gamma_{\pm }&=2\Gamma\left[ 1+\cos(2\theta_{0}) \right](1\pm \cos\theta_{0}).\label{GammaBpm}
\end{align}
\end{subequations}
Equation~(\ref{GammaBpm}) indicates that, to obtain the Fano line shapes for the reflection spectra, we need to ensure the condition $\Gamma_{\pm}\gg\Gamma_{\mp}$. Then the reflection spectrum near $\Lambda_{\pm}$ can be approximated by the Fano line shape characterized by Eq.~(\ref{Fano lineshape}). In the region of $\theta_{0}\in(0,\pi/2)\cup(\pi,3\pi/2)$, the phase shift should be chosen as either $\theta_{0}\in(0,0.1\pi)$ to ensure $\Gamma_{+}\gg\Gamma_{-}$ or $\theta_{0}\in(\pi,1.1\pi)$ to ensure $\Gamma_{-}\gg\Gamma_{+}$. Therefore the reflection spectra in Fig.~\ref{RBvsDelta}(c) (with $\theta_{0}=0.05\pi$ yielding $\Gamma_{+}\gg\Gamma_{-}$) and~\ref{RBvsDelta}(g) (with $\theta_{0}=1.05\pi$ leading to $\Gamma_{-}\gg\Gamma_{+}$) behave as the Fano line shapes. In addition, when $\theta_{0}$ takes these typical values, we can find that there will be a minimum for the reflection coefficient which occurs at $\Delta=-2\Gamma\tan\theta_{0}-g/\cos\theta_{0}$, which depends on the coupling strength $g$ and the phase shift $\theta_{0}$.
\begin{figure}[tbp]
\center\includegraphics[width=0.48\textwidth]{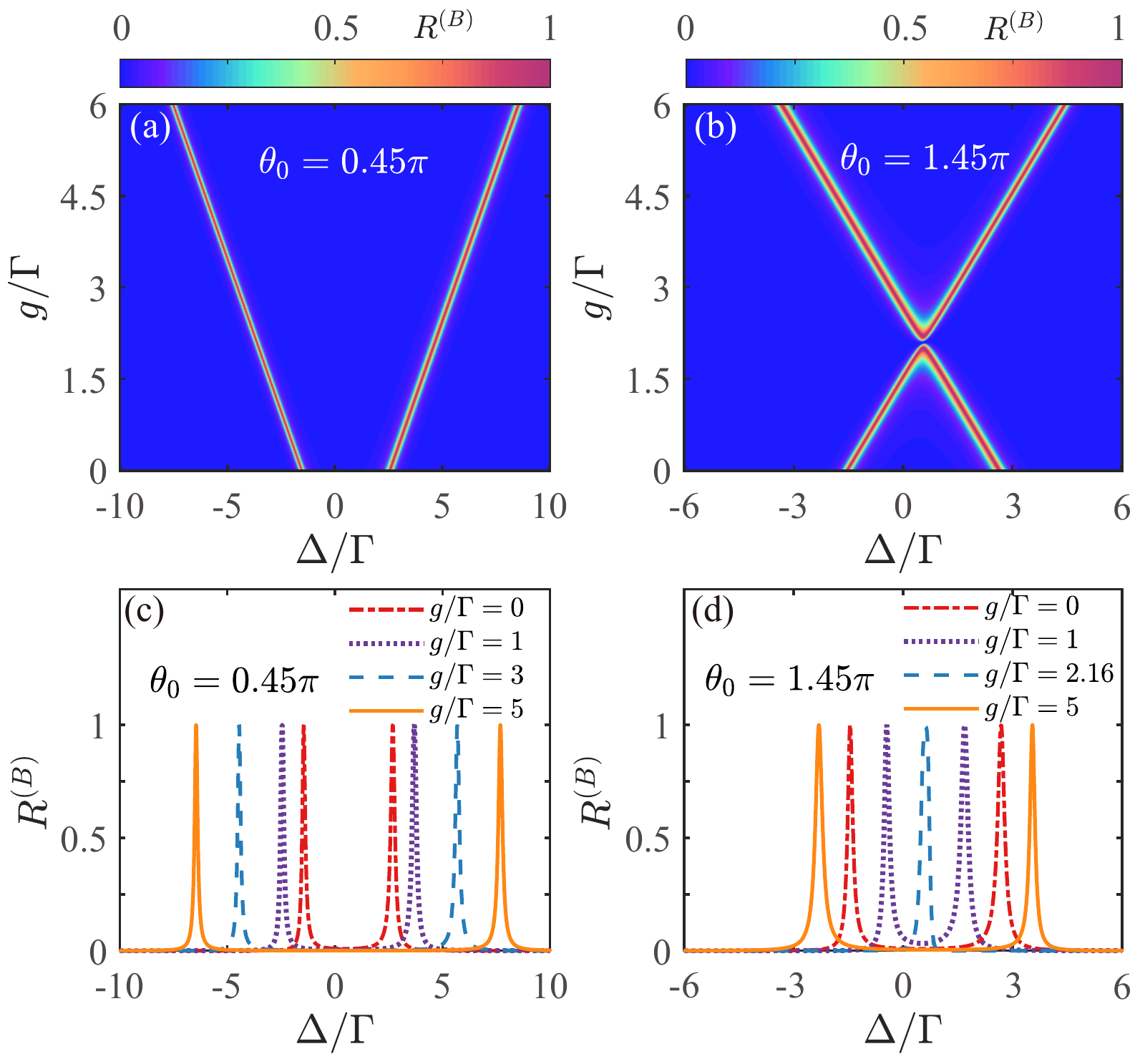}
\caption{(a) and (b) Reflection coefficient $R^{(B)}$ as a function of $\Delta$ and $g$ with $\theta_{0}=0.45\pi$ and $1.45\pi$, respectively. The curves in (c) and (d) show the profiles of (a) and (b) at given values of $g$.}
\label{RBvsDeltaandg}
\end{figure}

Specifically, when $\theta_{0}=(n+1/2)\pi$ with an integer $n$,  the reflection peak disappears completely, because the artificial giant molecule is decoupled from the waveguide. Thus the input photon can transmit the waveguide completely as shown in Figs.~\ref{RBvsDelta}(e) and~\ref{RBvsDelta}(i). However, when $\theta_{0}$ approaches to $\theta_{0}=(n+1/2)\pi$ such as $\theta_{0}=0.45\pi$ and $1.45\pi$ [see Figs.~\ref{RBvsDelta}(d) and~\ref{RBvsDelta}(h)], the spectra exhibit the vacuum Rabi-splitting-like line shapes. To explain this phenomenon, we choose $\theta_{0}=\pi/2+\delta$ and $3\pi/2+\delta$ with $|\delta|\ll1$, and then the transmission and reflection amplitudes can be approximated as
\begin{subequations}
\begin{eqnarray}
t^{(B)} &\approx &\frac{-\left(\Delta+4\Gamma\delta \right)^{2}+\left(g\pm2\Gamma\right)^{2}}{i\left(\Delta+4\Gamma\delta\right)\left[
i\left(\Delta+4\Gamma\delta\right)-8\Gamma\delta^{2}\right]+\left(g\pm2\Gamma\right)^{2}},\label{rabisplitt}\\
r^{(B)}&\approx &\frac{8\Gamma\left(g\pm2\Gamma\right) \delta^{2}}{i\left(\Delta+4\Gamma\delta\right)\left[i\left(\Delta+4\Gamma\delta\right)
-8\Gamma\delta^{2}\right]+\left(g\pm2\Gamma\right)^{2}}.\label{rabisplitr}
\end{eqnarray}
\end{subequations}
Equations~(\ref{rabisplitt}) and~(\ref{rabisplitr}) show that we can obtain the two peaks of the vacuum Rabi splitting spectra, which are located at $\Delta=-4\Gamma\delta\pm\left(g+2\Gamma\right)$ for $\theta_{0}=\pi/2+\delta$  and $\Delta=-4\Gamma\delta\pm\left(g-2\Gamma\right)$ for $\theta_{0}=3\pi/2+\delta$. The corresponding separations between the two peaks in the two cases are given by $d^{(B)}=2(g+2\Gamma)$ and $|2(g-2\Gamma)|$, respectively. Therefore, we see that the separation $d^{(B)}$ between the two peaks in Fig.~\ref{RBvsDelta}(d) is larger than that in Fig.~\ref{RBvsDelta}(h).  This feature is determined by the joint influence of the direction coupling $g$ and the exchange interaction $g_{ab}=\Gamma[3\sin\theta_{0}+\sin(3\theta_{0})]$  for the two braided giant atoms~\cite{Kockum2018}. When $\theta_{0}=(n+1/2)\pi$, the individual decays for the two giant atoms are zero, whereas the exchange interaction between them is nonzero, which is also called the decoherence-free interaction~\cite{Kockum2018}. Therefore, by choosing $\theta_{0}=\pi/2+\delta$, we can obtain $g_{ab}\simeq2\Gamma$, leading to a positive contribution $2g_{ab}=4\Gamma$ to the separation $d^{(B)}$ between the two reflection peaks~\cite{Cai2021}. However, when $\theta_{0}=3\pi/2+\delta$, the decoherence-free interaction is $g_{ab}\simeq-2\Gamma$, which results in a negative contribution $2g_{ab}=-4\Gamma$ to $d^{(B)}$.

Similar to the separated-coupling case, we also investigate the dependence of the reflection spectra $R^{(B)}$ on the coupling strength $g$. Figures~\ref{RBvsDeltaandg}(a) and~\ref{RBvsDeltaandg}(b) show the reflection coefficient $R^{(B)}$ as functions of $\Delta$ and $g$ when $\theta_{0}=0.45\pi$ and $\theta_{0}=1.45\pi$, respectively. In these cases, the reflection spectra exhibit vacuum Rabi-splitting-like line shapes [see Figs.~\ref{RBvsDelta}(d) and~\ref{RBvsDelta}(h)]. It can be seen from Fig.~\ref{RBvsDeltaandg}(a) that there are two reflection peaks for both $g=0$ and $g\neq 0$, which is different from the spectral feature in the separated-coupling case [see Fig.~\ref{RSvsDeltaandg}]. Based on the locations of the reflection coefficient $R^{(B)}$, the separation between the two reflection peaks is given by $d^{(B)}=2\sqrt{4\Gamma^{2}\left[1-\cos\theta_{0}\cos(3\theta_{0})\right] +g^{2}+2g\Gamma\left[3\sin\theta_{0}+\sin(3\theta_{0})\right]}$. Additionally, in terms of Eq.~(\ref{rabisplitr}), the separation between the peaks can be approximated as $d^{(B)}=2(g+\Gamma)$ at $\theta_{0}=0.45\pi$, which increases monotonically as $g$ increases. Note that the width of the two peaks is independent of the value of $g$ and the width of the right peak is slightly larger than that of the left one. However, the separation of the two reflection peaks exhibits a nonmonotonic behavior when $\theta_{0}=1.45\pi$, as shown in Fig.~\ref{RBvsDeltaandg}(b). To explain this feature, we substitute $\theta_{0}=1.45\pi$ into $d^{(B)}$ and find that it becomes an imaginary number when $g/\Gamma\in(1.975,2.169)$. Then in this region there is only one reflection peak appearing at $\Delta=2\Gamma\sin(2\theta_{0})$ [see the blue dashed line in Fig.~\ref{RBvsDeltaandg}(d)]. It can be seen from the profiles in Fig.~\ref{RBvsDeltaandg}(d) that, the distance between the two reflection peaks gradually decreases as $g$ goes from zero to $g/\Gamma=1.975$. As $g$ continues to increase from $g/\Gamma=2.169$, the separation between the two peaks begins to increase monotonously. Meanwhile, we can also obtain an approximate expression $d^{(B)}=|2(g-2\Gamma)|$ for the separation between the two peaks when $\theta_{0}=1.45\pi$ [corresponding to the case of $\theta_{0}=3\pi/2+\delta$]. This indicates that the separation $d^{(B)}$ between the two reflection peaks increases with the increase of $g$, only if the direct coupling strength $g$ is larger than the decoherence-free interaction $g_{ab}$ between the two giant atoms. Based on the above analyses, we know that the vacuum Rabi-splitting-like line shapes can be obtained in the braided-coupling case, in which the locations of the peaks are tunable by changing the coupling strength $g$ and the phase shift $\theta_{0}$.
\begin{figure*}[tbp]
\center\includegraphics[width=0.97\textwidth]{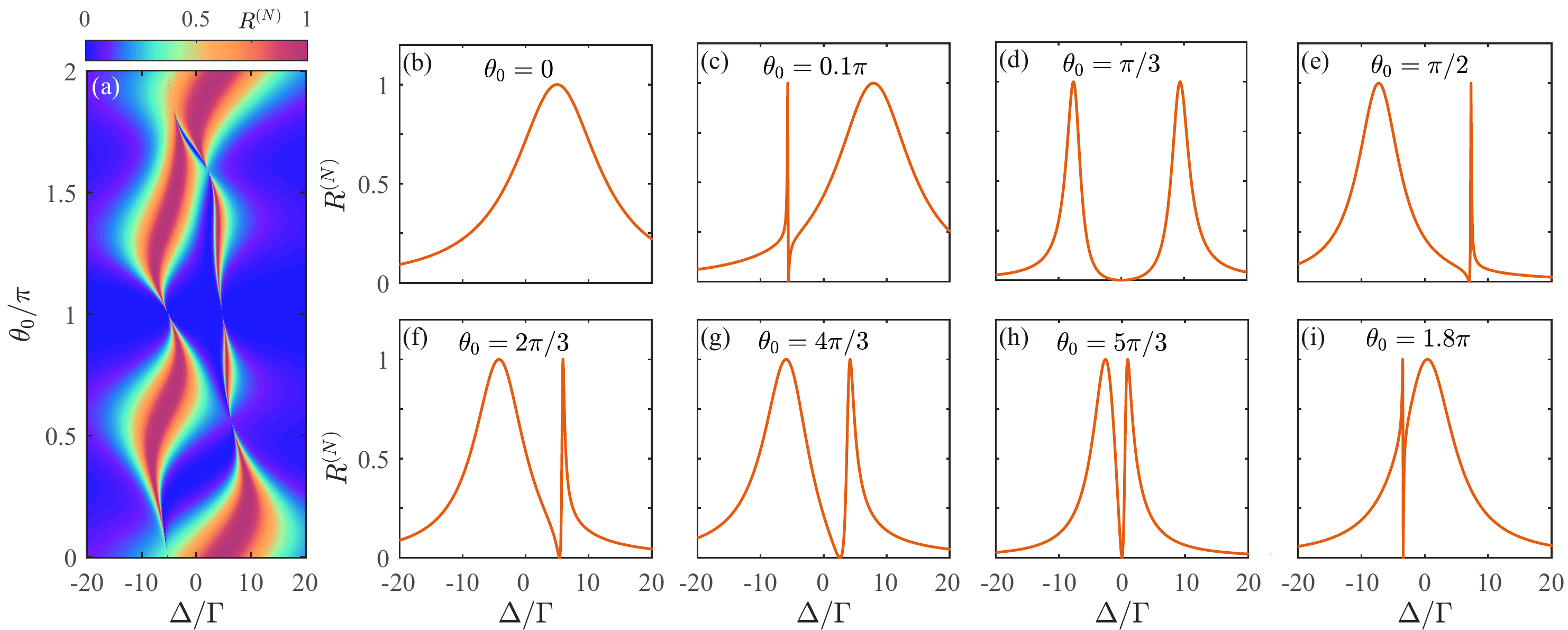}
\caption{(a) Reflection coefficient $R^{(N)}$ as a function of the detuning $\Delta$ and the phase shift $\theta_{0}$. The profiles of panel (a) are shown by the curves in (b)-(i) at different phases: (b) $\theta_{0}=0$, (c) $\theta_{0}=0.1\pi$, (d) $\theta_{0}=\pi/3$, (e) $\theta_{0}=\pi/2$, (f) $\theta_{0}=2\pi/3$, (g) $\theta_{0}=4\pi/3$, (h) $\theta_{0}=5\pi/3$, and (i) $\theta_{0}=1.8\pi$. In all panels, $g/\Gamma=5$ is taken.}
\label{RNvsDelta}
\end{figure*}

\subsection{Nested-coupling case}
We now turn to the nested-coupling case, as shown in Fig.~\ref{model_v1}(c).  In this case, the corresponding coefficients in $P_{i}(x)$ and $Q_{i}(x)$ are $m_{N}=3$, $n_{N}=1$, and $l_{N}=2$. Taking a similar method, the transmission and reflection amplitudes are obtained as
\begin{widetext}
\begin{subequations}
\begin{align}
t^{(N)}&=\frac{-\left[ \Delta -2\Gamma _{2}\sin\theta\right]
\left[ \Delta -2\Gamma _{1}\sin \left( 3\theta \right) \right]+\left \{ g+2\sqrt{\Gamma _{1}\Gamma _{2}}\left[ \sin \theta
+\sin \left(2\theta \right) \right] \right \}^{2}}{\left[ i\Delta -2\Gamma _{2}\left(1+e^{i\theta }\right) \right] \left[i\Delta -2\Gamma _{1}\left(1+e^{3i\theta }\right) \right]
-\left[ ig+2\sqrt{\Gamma _{1}\Gamma _{2}}e^{i\theta }\left( 1+e^{i\theta }\right) \right]^{2}},\label{tn}\\
r^{(N)}&=\frac{4ie^{3i\theta }\cos \left( \frac{\theta }{2}\right) ^{2}\left\{\Delta \left[ \left( 3\Gamma _{1}+\Gamma _{2}\right) -4\Gamma _{1}\cos
\theta +2\Gamma _{1}\cos \left( 2\theta \right) \right] +4\Gamma _{1}\Gamma_{2}\left[ \sin \left( 2\theta \right) -\sin\theta\right]
-2g\sqrt{\Gamma _{1}\Gamma _{2}}\left( 1-2\cos \theta \right) \right\} }{\left[ i\Delta -2\Gamma _{2}\left( 1+e^{i\theta }\right) \right] \left[i\Delta -2\Gamma _{1}\left(1+e^{3i\theta }\right) \right]-\left[ ig+2\sqrt{\Gamma _{1}\Gamma _{2}}e^{i\theta }\left( 1+e^{i\theta }\right) \right]^{2}}\label{rn}.
\end{align}
\end{subequations}
For simplicity, we consider the Markovian limit $\theta\approx\theta_{0}$ and the same decay rates $\Gamma_{1}=\Gamma_{2}=\Gamma$, and then $t^{(N)}$ and $r^{(N)}$ become
\begin{subequations}
\begin{eqnarray}
t^{(N)}&=&\frac{-\left[ \Delta -2\Gamma \sin\theta_{0}\right]
\left[ \Delta -2\Gamma \sin(3\theta_{0}) \right]+\left \{g+2\Gamma \left[\sin\theta_{0}
+\sin(2\theta_{0})\right]\right \}^{2}}{\left[ i\Delta -2\Gamma(1+e^{i\theta_{0}}) \right] \left[ i\Delta
-2\Gamma(1+e^{3i\theta_{0}})\right] -\left[ ig+2\Gamma e^{i\theta_{0}}(1+e^{i\theta_{0}}) \right] ^{2}}, \label{tnGammaeq}\\
r^{(N)}&=&\frac{8i\Gamma e^{3i\theta_{0}}\cos\left(\frac{\theta_{0}}{2}\right)
^{2}\left\{ \Delta \left[ 2-2\cos\theta_{0}+\cos(2\theta_{0})\right]
+2\Gamma \left[\sin(2\theta_{0}) -\sin\theta_{0})\right] -g(1-2\cos\theta_{0}) \right\}}{\left[ i\Delta -2\Gamma(1+e^{i\theta_{0}}) \right] \left[ i\Delta
-2\Gamma(1+e^{3i\theta_{0}})\right]-\left[ ig+2\Gamma e^{i\theta_{0}}(1+e^{i\theta_{0}}) \right] ^{2}}\label{rnGammaeq}.
\end{eqnarray}
\end{subequations}

In Fig.~\ref{RNvsDelta}(a), we display the reflection coefficient $R^{(N)}$ as a function of $\Delta$ and $\theta_{0}$. The dependence of $R^{(N)}$ on $\theta_{0}$ is also $2\pi$-periodic. In terms of Eqs.~(\ref{tnGammaeq}) and~(\ref{rnGammaeq}), we know that the reflection spectra have two peak values with $R^{(N)}=1$ under the condition $\Gamma ^{2}\left[\sin\theta_{0}-\sin(3\theta_{0})\right] ^{2}+\left\{ g+2\Gamma \left[\sin\theta_{0}+\sin(2\theta_{0})\right] \right\} ^{2}>0$, which are located at
\begin{equation}
\Delta =\Gamma \left[ \sin\theta_{0}+\sin(3\theta_{0}) \right]\pm\sqrt{\Gamma ^{2}\left[\sin\theta_{0}-\sin(3\theta_{0})\right] ^{2}+\left\{ g+2\Gamma \left[\sin\theta_{0}+\sin(2\theta_{0})\right] \right\} ^{2}},
\end{equation}
\end{widetext}
where the phase shift $\theta_{0}\neq2n\pi$ ($n$ is an integer). As $\theta_{0}$ increases, there will appear a reflection minimum $R^{(N)}=0$ at
\begin{equation}
\Delta =\frac{g( 1-2\cos\theta_{0}) -2\Gamma \left[ \sin(2\theta_{0}) -\sin\theta_{0}\right] }{2-2\cos\theta_{0}+\cos(2\theta_{0})},\label{minimumDeltaN}
\end{equation}
except for the phase shift $\theta_{0}=(2n+1)\pi$.
\begin{figure}[b]
\center\includegraphics[width=0.48\textwidth]{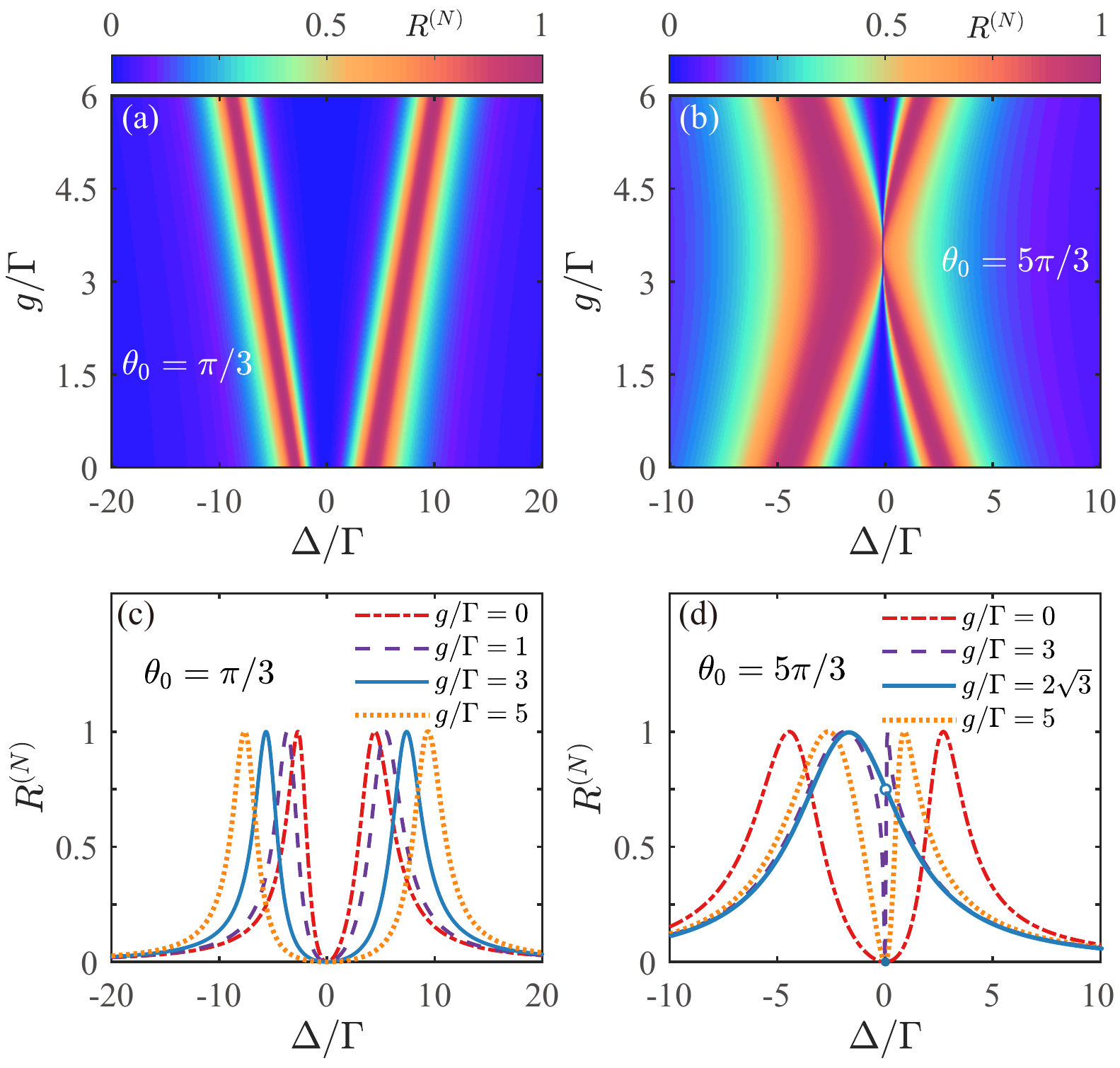}
\caption{(a) and (b) Reflection coefficient $R^{(N)}$ as a function of $\Delta$ and $g$ with $\theta_{0}=\pi/3$ and $5\pi/3$, respectively. The curves in (c) and (d) show the profiles of (a) and (b) at given values of $g$.}
\label{RNvsDeltaandg}
\end{figure}

Figures~\ref{RNvsDelta}(b)-\ref{RNvsDelta}(i) show the profiles of the reflection spectra when the phase shift $\theta_{0}$ takes different values in the region of $\theta_{0}\in[0,2\pi]$ [note that here we also have the relation $R^{(N)}(\Delta,\theta_{0})\neq R^{(N)}(-\Delta,2\pi-\theta_{0})$]. As shown in Fig.~\ref{RNvsDelta}(b), when $\theta_{0}=0$, the reflection amplitude $r^{(N)}$ is reduced to $r^{(N)}=-8i\Gamma /\left(\Delta -g-8i\Gamma \right)$, which means that the reflection spectrum is a Lorentzian line shape centered at $\Delta=g$ with spectrum linewidth $16\Gamma$. In particular, according to Eq.~(\ref{minimumDeltaN}) we find that the minimum of the reflection appears at $\Delta=0$ when $\theta_{0}=\pi/3$ and $5\pi/3$. As shown in Figs.~\ref{RNvsDelta}(c)-\ref{RNvsDelta}(i), there will appear two reflection peaks with $R^{(N)}=1$ and one dip with $R^{(N)}=0$ when $\theta_{0}$ increases from zero to $\pi$ or from $\pi$ to $2\pi$. In particular, the reflection dip exists always between the two reflection peaks, which shows a different feature from the braided-coupling case, where the two reflection peaks always appear on the same side of the reflection minimum. On the other hand, from Figs.~\ref{RNvsDelta}(c) and~\ref{RNvsDelta}(i) we find that the reflection minimum appears at the left-hand side of $\Delta=0$ when $\theta_{0}\in(0,\pi/3)\cup(5\pi/3,2\pi)$. From Figs.~\ref{RNvsDelta}(e)-\ref{RNvsDelta}(g) we see that, when $\theta_{0}\in(\pi/3,\pi)\cup(\pi,5\pi/3)$, the reflection minimum appears at the right-hand side of $\Delta=0$. When $\theta_{0}=0.1\pi$, $\pi/2$, $2\pi/3$, and $1.8\pi$, the sharp Fano-like line shapes can be observed, as shown in  Figs.~\ref{RNvsDelta}(c),~\ref{RNvsDelta}(e),~\ref{RNvsDelta}(f), and~\ref{RNvsDelta}(i).
\begin{figure}[tbp]
\center\includegraphics[width=0.48\textwidth]{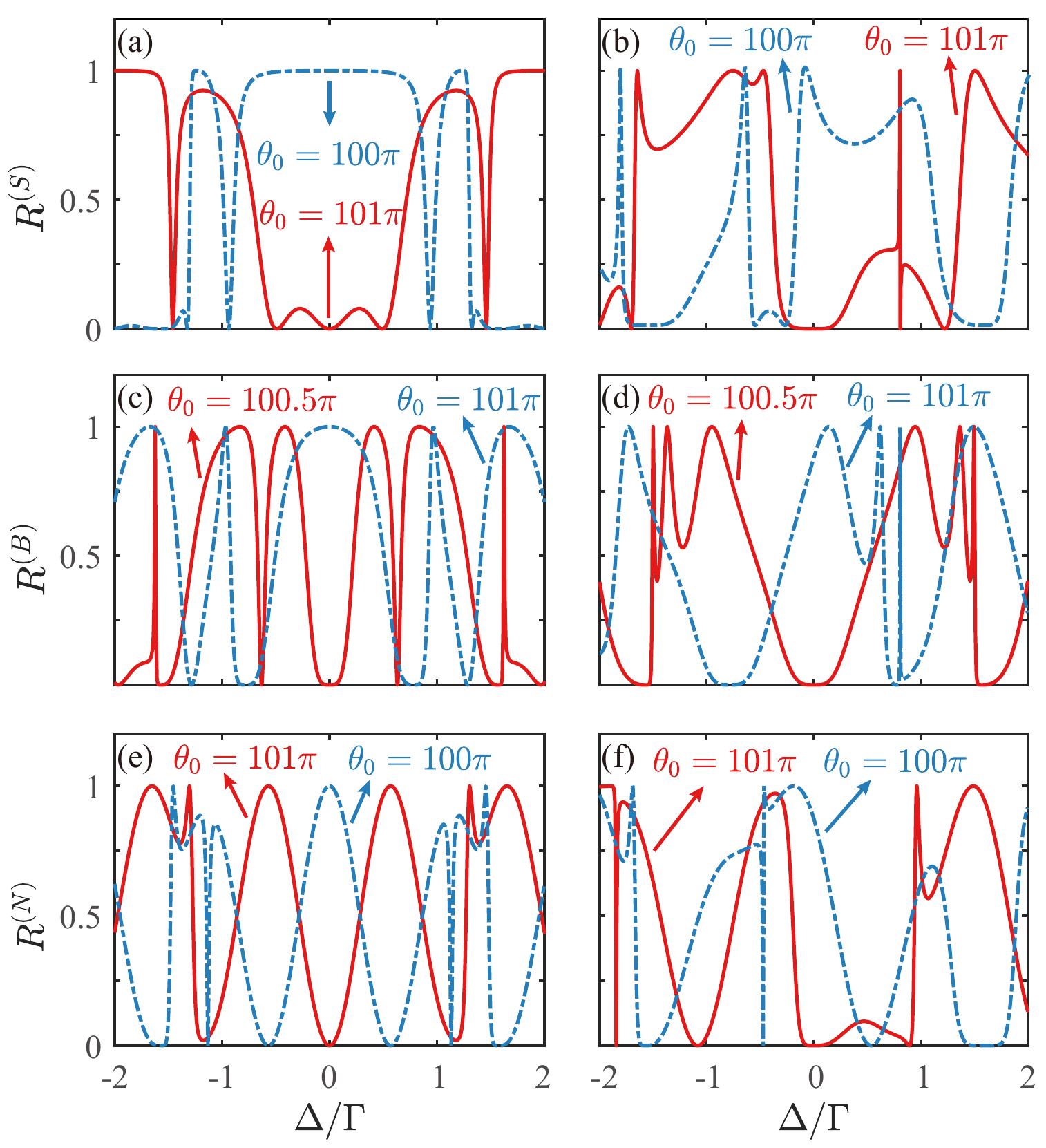}
\caption{Reflection coefficients (a) $R^{(S)}$,  (c) $R^{(B)}$,  and (e) $R^{(N)}$ as functions of the detuning $\Delta$ when $g/\Gamma=0$. Reflection coefficients (b) $R^{(S)}$,  (d) $R^{(B)}$,  and (f) $R^{(N)}$ as functions of the detuning $\Delta$ when $g/\Gamma=3$. In all panels, we choose $\tau\Gamma=2$.}
\label{RvsDelta}
\end{figure}

Similarly, we also investigate the dependence of the reflection coefficient $R^{(N)}$ on the coupling strength $g$. Figures~\ref{RNvsDeltaandg}(a) and~\ref{RNvsDeltaandg}(b) show the reflection coefficient as a function of $\Delta$ and $g$ when $\theta_{0}=\pi/3$ and $\theta_{0}=5\pi/3$, respectively. In the cases for the two specific phase shifts, the minimum of the reflection is located at $\Delta=0$. As shown in Fig.~\ref{RNvsDeltaandg}(a), the separation between the two reflection peaks gradually increases with the increase of $g$. This feature can be explained from the separation between the two reflection peaks $d^{(N)}=2\sqrt{\Gamma^{2}\left[\sin\theta_{0}-\sin(3\theta_{0})\right] ^{2}+\left\{g+2\Gamma\left[\sin\theta_{0}+\sin(2\theta_{0})\right]\right\}^{2}}$ in the case of $\theta_{0}=\pi/3$, which is always a monotonically increasing function of $g$. However, from Fig.~\ref{RNvsDeltaandg}(b) it can be found that the two reflection peaks first gradually approach until a minimum is reached, and then gradually separate when $\theta_{0}=5\pi/3$. This is because $d^{(N)}$ becomes a nonmonotonic function of $g$ in the case of $\theta_{0}=5\pi/3$. In Figs.~\ref{RNvsDeltaandg}(c) and~\ref{RNvsDeltaandg}(d), we plot the profiles of the reflection spectra at some given coupling strength $g$, which show that the location of the reflection minimum $R^{(N)}=0$ is immune to the changing of $g$ when $\theta_{0}=\pi/3$ and $\theta_{0}=5\pi/3$. To understand this phenomenon, we substitute $\theta_{0}=\pi/3$ and $5\pi/3$ into Eq.~(\ref{minimumDeltaN}) and can find that the reflection minimum $R^{(N)}=0$ is always located at $\Delta=0$, which is independent of $g$. In addition, the profiles of $R^{(N)}$ in Fig.~\ref{RNvsDeltaandg}(d) at $\theta_{0}=5\pi/3$ indicate that when $g/\Gamma$ goes from $0$ to $2\sqrt 3$, the two peaks are getting closer and closer. When $g/\Gamma$ approaches to $2\sqrt 3$, such as $g/\Gamma=3$, the reflection spectrum exhibits a sharp Fano-like line shape [see the purple dashed line in Fig.~\ref{RNvsDeltaandg}(d)]. Specifically, there is only one peak and one point with $R^{(N)}=0$ located at $\Delta=0$ when $g/\Gamma=2\sqrt 3$. As $g$ increases from $g/\Gamma=2\sqrt 3$ to larger values, the separation between the two peaks gradually increases.

\section{Single-photon scattering in the Non-Markovian regime}\label{NonMarkovian}
In this section, we study the single-photon scattering in the non-Markovian regime, in which the retarded effect induced by the $\tau$ term in Eq.~(\ref{theta}) cannot be neglected. This non-Markovian effect should be taken into account in some physical systems, such as a superconducting qubit coupled to surface acoustic waves~\cite{Guo2017,Andersson2019}.

In this non-Markovian regime, the accumulated phase shift is $\theta=\tau\Delta+\theta_{0}$. As a result, the phase shift $\theta_{0}$ in Eqs.~(\ref{tsGammaeq}) and~(\ref{rsGammaeq}),~(\ref{tbGammaeq}) and~(\ref{rbGammaeq}), and~(\ref{tnGammaeq}) and~(\ref{rnGammaeq}) should be replaced by $\theta$. To see the influence of the non-Markovian effect on the reflection coefficients in these three coupling configurations, we first consider the case of $g=0$, which corresponds to the system with two independent giant atoms coupled to a one-dimensional waveguide. It should be noted that even if $\theta_{0}$ is a periodic function of $2\pi$, we need to take $\theta_{0}=\Omega\tau\gg\Gamma\tau$ to meet the parameter condition $\Gamma/\Omega\ll1$ for the rotating-wave approximation. As shown in Figs.~\ref{RvsDelta}(a),~\ref{RvsDelta}(c), and~\ref{RvsDelta}(e), we plot the reflection coefficients $R^{(S)}$, $R^{(B)}$, and $R^{(N)}$ as functions of $\Delta$ at given values of $\tau\Gamma$ and $\theta_{0}$ when the coupling strength $g=0$. It can be seen from Figs.~\ref{RvsDelta}(a),~\ref{RvsDelta}(c), and~\ref{RvsDelta}(e) that, the reflection coefficients are symmetric to $\Delta=0$ and exhibit complicated structures and there appear more reflection peaks and dips, which is due to the dependence of the phase shift $\theta$ on the detuning $\Delta$. According to the analyses in Sec.~\ref{SPTAGM}, the locations of the complete reflection peaks for $R^{(S)}$, $R^{(B)}$, and $R^{(N)}$ in the case of $g=0$ are, respectively, determined by
\begin{eqnarray}
\Delta^{(S)}\!&=\!&2\Gamma\sin\theta^{(S)}, \notag\\
\Delta^{(B)}\!&=\!&2\Gamma\sin(2\theta^{(B)})\pm2\Gamma\sqrt{\left[1-\cos\theta^{(B)}\cos (3\theta^{(B)})\right]}, \notag\\
\Delta ^{(N)} &=&\Gamma \left[ \sin \theta ^{(N)}+\sin (3\theta ^{(N)})\right] \notag \\
&&\pm \Gamma \sqrt{\left[ \sin \theta ^{(N)}-\sin (3\theta ^{(N)})\right]^{2}+4\left[ \sin \theta ^{(N)}+\sin (2\theta ^{(N)})\right] ^{2}}, \notag\\
\end{eqnarray}
which are transcendental equations, where $\theta^{(S)}$, $\theta^{(B)}$, and $\theta^{(N)}$ are the phase in the three coupling configurations defined by $\theta^{(f)}=\tau\Delta^{(f)}+\theta_{0}$ for $f=S$, $B$, and $N$. Here, we use the superscript $S$, $B$, and $N$ to mark the three coupling configurations.  In the case of $g\neq0$, as shown in Figs.~\ref{RvsDelta}(b),~\ref{RvsDelta}(d), and~\ref{RvsDelta}(f), the reflection coefficients $R^{(S)}$, $R^{(B)}$, and $R^{(N)}$ also behave as complicated but asymmetric line shapes (except for the braided-coupling case when $\theta_{0}=100.5\pi$). Meanwhile, more reflection peaks and dips appear, and the locations of the complete reflection peaks for $R^{(S)}$, $R^{(B)}$, and $R^{(N)}$ are also determined by the transcendental equations
\begin{eqnarray}
\Delta^{(S)}&=&2\Gamma\sin \theta^{(S)}\pm \sqrt{G^{(S)}},\notag \\
\Delta^{(B)}&=&2\Gamma\sin (2\theta ^{(B)})\pm \sqrt{4\Gamma ^{2}\left[1-\cos \theta ^{(B)}\cos (3\theta ^{(B)})\right] +G^{(B)}},  \notag \\
\Delta^{(N)}&=&\Gamma\left[\sin\theta ^{(N)}+\sin (3\theta ^{(N)})\right]\notag \\
&&\pm \sqrt{\Gamma^{2}\left[\sin\theta^{(N)}-\sin (3\theta ^{(N)})\right]^{2}+G^{(N)}},
\end{eqnarray}
with
\begin{eqnarray}
G^{(S)} &=&g^{2}+4g\Gamma \sin (2\theta ^{(S)})(1+\cos \theta ^{(S)}),  \notag\\
G^{(B)} &=&g^{2}\!+\!2g\Gamma \left[ 3\sin \theta ^{(B)}+\sin (3\theta ^{(B)})\right] ,  \notag \\
G^{(N)} &=&\left\{ g+2\Gamma \left[ \sin \theta ^{(N)}+\sin (2\theta ^{(N)})\right] \right\}^{2}.
\end{eqnarray}
In addition, we can find that the reflection coefficients $R^{(S)}$, $R^{(B)}$, and $R^{(N)}$ are no longer characterized by the Lorentzian line shapes when $\theta_{0}=2n\pi$ in the cases of both $g=0$ and $g\neq0$, which is different from the situations in the Markovian regime. In the non-Markovian regime, meanwhile, when the phase shift $\theta_{0}=(2n+1)\pi$ for the separated- and nested-coupling cases and $\theta_{0}=(n+1/2)\pi$ for the braided-coupling case, the retarded effect induced by the propagating time $\tau$ leads to the revival of the vanished reflection spectra in the Markovian regime.
\begin{figure}[tbp]
\center\includegraphics[width=0.48\textwidth]{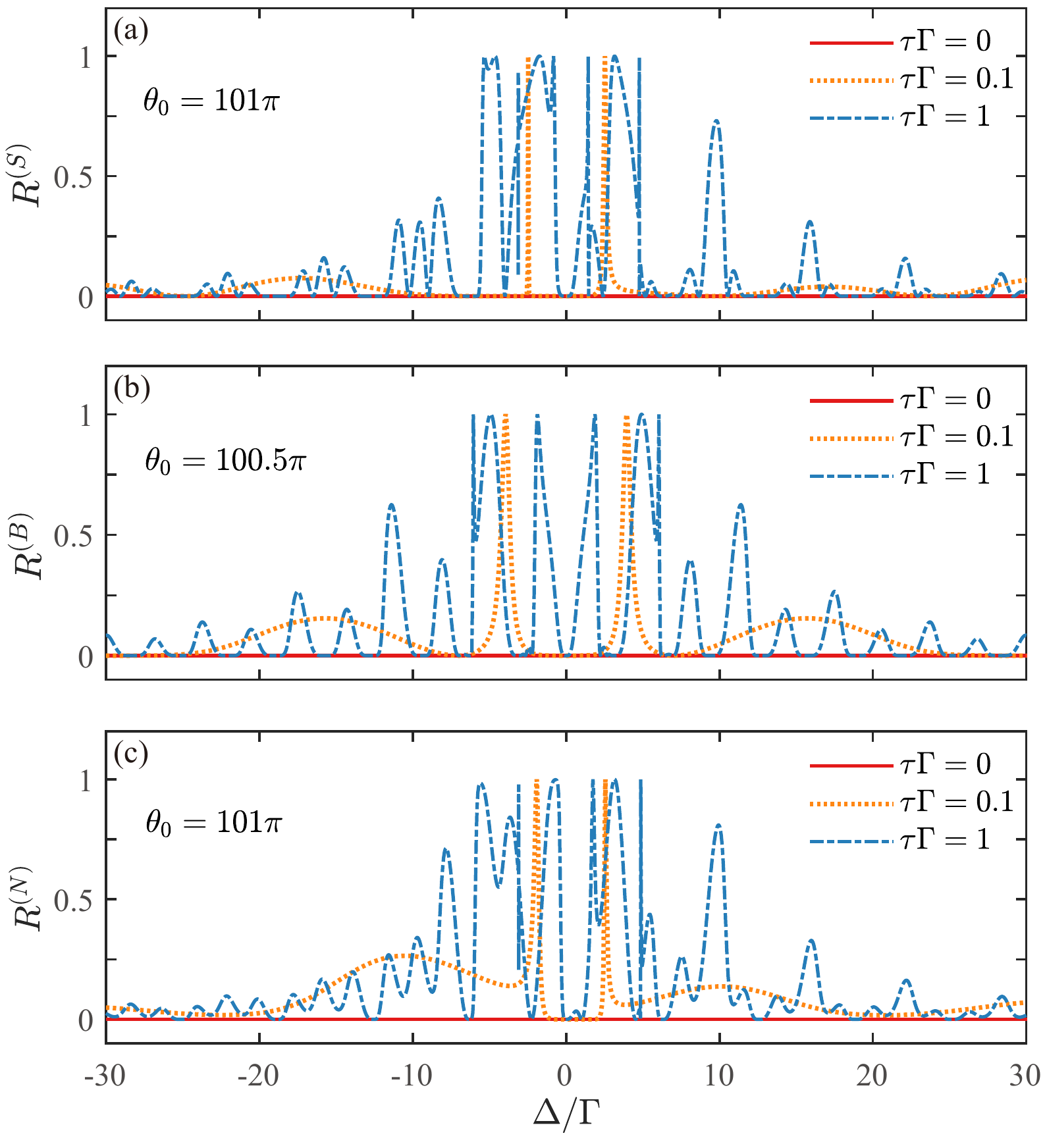}
\caption{Reflection coefficients (a) $R^{(S)}$, (b) $R^{(B)}$, and (c) $R^{(N)}$ vs the detuning $\Delta$ at different values of $\tau$ and $\theta_{0}$. In panels (a), (b), and (c), the phase shifts are $\theta_{0}=101\pi$, $100.5\pi$, and $101\pi$, respectively. The other parameter is $g/\Gamma=3$.}
\label{Rvstau}
\end{figure}

To better understand the single-photon scattering of the three coupling configurations in the non-Markovian regime, we plot $R^{(S)}$, $R^{(B)}$, and $R^{(N)}$ as functions of $\Delta$ when $\tau\Gamma=0.001$, $0.1$, and $1$ in Fig.~\ref{Rvstau}. It can be seen that in the Markovian regime $\tau\Gamma=0.001$ the single-photon is completely transmitted in all three coupling cases when $\theta_{0}=101\pi$ for $R^{(S)}$ and $R^{(N)}$, and $\theta_{0}=100.5\pi$ for $R^{(B)}$. In the non-Markovian regime, as the propagating time $\tau$ increases, the oscillation of the reflection spectra and their sensitivity to the detuning $\Delta$ are enhanced. Another interesting feature is the revival of the multiple reflection peaks in the non-Markovian regime. As shown in Figs.~\ref{Rvstau}(a) and~\ref{Rvstau}(c), the reflection spectra exhibit three-peak structures away from $\Delta=\pm g$. However, different from the separated- and nested-coupling cases, there are no three-peak structures for the reflection spectra $R^{(B)}$ in Fig.~\ref{Rvstau}(b), but we can see that the reflection spectra $R^{(B)}$ are always symmetric to $\Delta=0$ due to the relation $R^{(B)}(\Delta,\tau\Delta+\theta_{0})=R^{(B)}(-\Delta,-\tau\Delta+\theta_{0})$ at $\theta_{0}=(n+1/2)\pi$. In addition, the maximal values of the reflection coefficients $R^{(S)}$, $R^{(B)}$, and $R^{(N)}$ in the non-Markovian regime gradually decrease with the increase of the detuning $\Delta$.
\section{Discussion and Conclusion}\label{Conclusions}
Finally, we present some discussions on the experimental implementation of the single-photon scattering in the giant-molecule waveguide-QED system. In this system, there exist both the two-point coupling of each giant atom with the waveguide and the inner coupling of the two giant atoms. Therefore, the candidate experimental setups should be able to realize the two kinds of couplings. Concretely, the experimental demonstration of the giant atoms has been realized by coupling transmon qubits to surface acoustic waves~\cite{Gustafsson2014,Manenti2017,Andersson2019} or coupling Xmon qubits to a meandering transmission line~\cite{Kannan2020,Vadiraj2021}. By changing the frequencies of the qubits, the accumulated phase shift $\theta_{0}$ between neighboring coupling points can be tuned. The inner coupling between the two transmon qubits can be realized by connecting them to a common and tunable inductor. In our simulations, we take the inner coupling strength $g/\Gamma>1$. It has been reported that the inner coupling strength can be adjusted to reach tens of MHz~\cite{Chen2014}. In addition, the recent experiments have shown that the coupling strength between the transmon qubits and the transmission lines can reach MHz~\cite{Schoelkopf2010,Peng2018}. For the single-photon scattering in the non-Markovian regime of this scheme,  we consider that the propagating time $\tau$ of single photons between two neighboring coupling points is comparable to the lifetime $1/\Gamma$ of the transmon qubits, i.e., $\tau\Gamma\sim1$. In Ref.~\cite{Andersson2019}, the giant atom is achieved by coupling a transmon qubit with multiple interdigital transducers to SAWs. As the speed of SAWs is very slow, the parameter condition $\tau\Gamma\sim1$ can be well satisfied. Moreover, we can enlarge the distances between the neighboring coupling points, and hence the system can work in the non-Markovian regime~\cite{Longhi2020}. All these advances indicate that the giant-molecule waveguide-QED system proposed in our paper can be implemented with current and near-future conditions.

In conclusion, we have studied single-photon transport in a one-dimensional waveguide coupled to a giant artificial molecule consisting of two coupled giant atoms. The system has three different coupling configurations based on the coupling points of the two giant atoms with the waveguide. In particular, we have considered the single-photon scattering in both the Markovian and non-Markovian regimes, in which the retarded effect of a single photon propagating between two coupling points can and cannot be neglected, respectively. We have obtained the exact results of the single-photon scattering coefficients, and have found that the phase delay, the coupling configurations, and the coupling strength between the two giant atoms determine the scattering properties of the single-photon transport. The scattering spectra and spectral features have been analyzed in detail. The single-photon transport can be controlled perfectly and hence single-photon switch can be realized in this system.  This paper will pave the way for the study of single-photon quantum devices in giant-molecule waveguide-QED systems.

\begin{acknowledgments}
The authors thank Xun-Wei Xu and Ye-Xiong Zeng for helpful discussions. J.-Q.L. was supported in part by National Natural Science Foundation of China (Grants No.~12175061, No.~11822501, No.~11774087, and No.~11935006), the Science and Technology Innovation Program of Hunan Province (Grants No.~2021RC4029 and No.~2020RC4047), and Hunan Science and Technology Plan Project (Grant No.~2017XK2018). J.-F.H. is supported in part by the National Natural Science Foundation of China (Grant No.~12075083), and Natural Science Foundation of Hunan Province, China (Grant No.~2020JJ5345).
\end{acknowledgments}

\end{document}